\documentclass[a4paper,11pt]{article}
\usepackage{theorem}
\usepackage{titlesec}
\usepackage[a4paper,height=240mm,width=162mm]{geometry}
\usepackage[matrix,arrow,curve,cmtip]{xy}
\usepackage{amssymb}
\usepackage{latexsym}
\usepackage{epic}

\titleformat{\section}[hang]%
{\bfseries\filcenter\large}{\thesection.}{1ex}{}%

\def\dot{\makebox(0,0){$\bullet$}}

\def\lett#1{\makebox(0,0){${#1}$}}



\def\to{\mbox{$\xymatrix@1@C=5mm{\ar@{->}[r]&}$}}
\def\shorterto{\mbox{$\xymatrix@1@C=4.5mm{\ar@{->}[r]&}$}}
\def\tto{\mbox{$\xymatrix@1@C=5mm{\ar@{=>}[r]&}$}}
\def\isosign{\begin{picture}(0,0)\put(-1.5,0.3){$\sim$}\end{picture}}
\def\isosignn{\begin{picture}(0,0)\put(-1.75,0.5){$\sim$}\end{picture}}
\def\iso{\mbox{$\xymatrix@1@C=6mm{\ar@{->}[r]^{\isosign}&}$}}
\def\iiso{\mbox{$\xymatrix@1@C=7mm{\ar@{=>}[r]^{\isosignn}&}$}}
\def\parto{\mbox{$\xymatrix@1@C=5mm{\ar@{-->}[r]&}$}}

\newtheorem{theorem}{Theorem}[section]
\newtheorem{lemma}[theorem]{Lemma}
{\theorembodyfont{\upshape}\newtheorem{definition}[theorem]{Definition}}
\newtheorem{proposition}[theorem]{Proposition}
\newtheorem{corollary}[theorem]{Corollary}
{\theorembodyfont{\upshape}\newtheorem{example}[theorem]{Example}} 
{\theorembodyfont{\upshape}}
\newenvironment{proof}{\begin{trivlist}\item[]{\bf Proof.}}{\hspace*{\fill} 
$\Box$ \end{trivlist}}
\newenvironment{sketchofproof}{\begin{trivlist}\item[]{\bf Sketch of the proof.}}{\hspace*{\fill} 
$\Box$ \end{trivlist}}
{\theorembodyfont{\upshape}\newtheorem{mycomment}[theorem]{\hspace{-1mm}}}

\def\:{\colon}

\def\L{\mathcal{L}}
\def\C{\mathcal{C}}
\renewcommand\P{\mathcal{P}}
\def\G{\mathcal{G}}
\def\ortho{^{\perp}}
\def\dblortho{^{\perp\perp}}
\def\c{^{\mathsf{c}}}
\def\*{\star}
\def\ParSet{\mathsf{ParSet}}
\def\ProjGeom{\mathsf{ProjGeom}}
\def\SemiVec{\mathsf{Vec}}
\def\Vec{\mathsf{Vec}}
\def\ProjLat{\mathsf{ProjLat}}
\def\HilbGeom{\mathsf{HilbGeom}}
\def\HilbLat{\mathsf{HilbLat}}
\def\PropSys{\mathsf{PropSys}}
\def\ContSemiHilb{\mathsf{GenHilb}}
\def\GenHilb{\mathsf{GenHilb}}
\def\dim{\mathsf{dim}}
\def\ker{\mathsf{ker}}
\def\im{\mathsf{im}}
\def\card{\mathsf{card}}

\def\Z{\mathcal{Z}}
\def\cl{\mathsf{cl}}
\def\f{_{\mathsf{f}}}
\def\inprod#1{\langle{#1}\rangle}
\def\RR{\mathbb{R}}
\def\CC{\mathbb{C}}
\def\HH{\mathbb{H}}
\def\iff{\Leftrightarrow}
\def\id{\mathsf{id}}
\def\cov{\lessdot}
\def\op{^{\mathsf{op}}}
\def\defeq{\mathrel{\mathop:}=}
\def\clvee{\triangledown}
\def\clbigvee{\mbox{
\begin{picture}(2.5,1)
\thicklines \put(-0.8,0){$\bigvee$} \put(-0.5,2.5){\line(1,0){2.3}}
\end{picture}}}
\def\rk{\mathsf{rk}}
\def\disju{\uplus}
\renewcommand{\phi}{\varphi}

\def\c{^{\mathsf{c}}}
\def\asterix{$^{\hspace{-0.4mm}\dagger}$\hspace{-0.7mm}}

\title{Propositional systems, Hilbert lattices \\ and generalized 
Hilbert spaces} 

\author{Isar Stubbe\footnote{%
Postdoctoral Fellow of the Research Foundation Flanders (FWO), 
Departement Wiskunde en Informatica,
Universiteit Antwerpen. 
E-mail: {\tt isar.stubbe@ua.ac.be}}
\ and Bart Van Steirteghem\footnote{%
Partially supported by FCT/POCTI/FEDER, 
Departamento de Matem\'atica,
Instituto Superior T\'ecnico, Lisboa. 
E-mail: {\tt bvans@math.ist.utl.pt}}}

\date{February 21, 2006\footnote{Final version published as: [I. Stubbe and B. Van Steirteghem, 2007] ``Propositional systems, Hilbert lattices and generalized Hilbert spaces'', Handbook of Quantum Logic and Quantum Structures: Quantum Structures (Eds.\ K. Engesser, D. M. Gabbay and D. Lehmann), Elsevier, pp.~477--524.}}

\begin{document}
\maketitle
\begin{quote} 
{\bf Abstract.} With this chapter we provide a compact yet complete 
survey of two most remarkable ``representation theorems'': every 
arguesian projective geometry is represented by an essentially 
unique vector space, and every arguesian Hilbert geometry is 
represented by an essentially unique generalized Hilbert space. 
C.~Piron's original representation theorem for propositional systems 
is then a corollary: it says that every irreducible, complete, 
atomistic, orthomodular lattice satisfying the covering law and of 
rank at least 4 is isomorphic to the lattice of closed subspaces of 
an essentially unique generalized Hilbert space. Piron's theorem 
combines abstract projective geometry with lattice theory. In fact, 
throughout this chapter we present the basic lattice theoretic 
aspects of abstract projective geometry: we prove the categorical 
equivalence of projective geometries and projective lattices, and 
the triple categorical equivalence of Hilbert geometries, Hilbert 
lattices and propositional systems.\\[2mm]
\end{quote}

\section{Introduction}\label{A}
\setcounter{theorem}{0}
{\bf Description of the problem.} The definition of a Hilbert space 
$H$ is all about a perfect marriage between linear algebra and 
topology: $H$ is a vector space together with an inner product such 
that the norm associated to the inner product turns $H$ into a 
complete metric space. As is well-known for any vector space, the 
one-dimensional linear subspaces of $H$ are the points of a {\it 
projective geometry}, the collinearity relation being coplanarity. 
In other words, the set $\L(H)$ of linear subspaces, ordered by 
inclusion, forms a so-called {\it projective lattice}.
\par
Using the metric topology on $H$ we can distinguish, amongst all 
linear subspaces, the closed ones: we will note the set of these as 
$\C(H)$. In fact, the inner product on $H$ induces an {\it 
orthogonality operator} on $\L(H)$ making it a {\it Hilbert 
lattice}, and the map
$$(\hspace{1ex})\dblortho\:\L(H)\to \L(H)\: A\mapsto A\dblortho$$
is a closure operator on $\L(H)$ whose fixpoints are precisely the 
elements of $\C(H)$. For many reasons, explained in detail elsewhere 
in this volume, it is the substructure $\C(H)\subseteq\L(H)$ -- and 
not $\L(H)$ itself -- which plays an important r\^ole in quantum 
logic; it is called a {\it propositional system}. 
\par
\begin{figure}
$$\xymatrix@=15mm{
\ContSemiHilb\ar[r]\ar[d] & \HilbGeom\ar[d]\ar@{<->}[r]^{\ \ \sim} & \HilbLat\ar@{<->}[r]^{\sim}\ar[d] & \PropSys \\
\SemiVec\ar[r] & \ProjGeom\ar@{<->}[r]^{\ \ \sim}
& \ProjLat }$$ \caption{A diagrammatic summary} \label{0}
\end{figure}
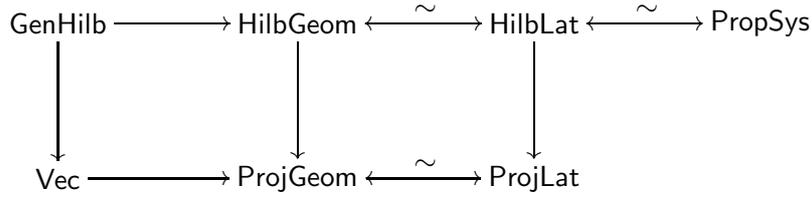

In this survey paper we wish to explain the lattice theoretic 
axiomatization of such a propositional system: we study necessary 
and sufficient conditions for an ordered set $(C,\leq)$ to be 
isomorphic to $(\C(H),\subseteq)$ for some (real, complex, 
quaternionic or generalized) Hilbert space $H$. As the above 
presentation suggests, this matter is intertwined with some deep 
results on projective geometry.
\\[2mm] 
{\bf Overview of contents.} Section \ref{B} of this 
paper presents the relevant definitions of, and some basic results 
on, abstract (also called `modern' or `synthetic') projective 
geometry. Following Cl.-A. Faure and A. Fr\"olicher's [2000] 
reference on the subject, we define a `projective geometry' as a set 
together with a ternary collinearity relation (satisfying suitable 
axioms). The one-dimensional subspaces of a vector space are an 
example of such a projective geometry, with coplanarity as the ternary 
relation. After discovering some particular properties of the 
ordered set of `subspaces' of such a projective geometry, we make an 
abstraction of this ordered set and call it a `projective lattice'. 
We then speak of `morphisms' between projective geometries, resp.\ 
projective lattices, and show that the category $\ProjGeom$ of 
projective geometries and the category $\ProjLat$ of projective 
lattices are equivalent. Vector spaces and `semilinear maps' form a 
third important category $\Vec$, and there is a functor 
$\Vec\to\ProjGeom$. The bottom row in figure \ref{0} summarizes this. 
\par
A projective geometry for which every line contains at least three 
points, is said to be `irreducible'. We deal with these in section \ref{C}, for this geometric fact has an important 
categorical significance [Faure and Fr\"olicher, 2000]: a projective geometry is irreducible 
precisely when it is not a non-trivial coproduct in $\ProjGeom$, and 
every projective geometry is the coproduct of irreducible ones. By 
the categorical equivalence between $\ProjGeom$ and $\ProjLat$, the 
``same'' result holds for projective lattices. The projective 
geometries in the image of the functor $\Vec\to\ProjGeom$ are always 
irreducible.
\par
Having set the scene, we deal in section \ref{D} with the linear 
representation of projective geometries (of dimension at least 2) 
and their morphisms, i.e.\ those objects and morphisms that lie in 
the image of the functor $\Vec\to\ProjGeom$. The First Fundamental 
Theorem, which is by now part of mathematical folklore, says that 
precisely the `arguesian' geometries (which include all geometries 
of dimension at least $3$) are ``linearizable''. The Second 
Fundamental Theorem characterizes the ``linearizable'' morphisms. 
[Holland, 1995, \S3] and [Faure, 2002] have some comments on the 
history of these results. We outline the proof of the First 
Fundamental Theorem as given in [Beutelspacher and Rosenbaum, 1998]; 
for a short proof of the Second Fundamental Theorem we refer to 
[Faure, 2002].
\par
Again following [Faure and Fr\"olicher, 2000], we turn in section 
\ref{E} to projective geometries that come with a binary 
orthogonality relation which satisfies certain axioms: so-called 
`Hilbert geometries'. The key example is given by the projective 
geometry of one-dimensional subspaces of a `generalized Hilbert 
space' (a notion due to C. Piron [1976]), with the orthogonality 
induced by the inner product. The projective lattice of subspaces of 
such a Hilbert geometry inherits an orthogonality operator which 
satisfies some specific conditions, and this leads to the notion of 
`Hilbert lattice'. The elements of a Hilbert lattice that equal 
their biorthogonal are said to be `(biorthogonally) closed'; they 
form a `propositional system' [Piron, 1976]: a complete, atomistic, 
orthomodular lattice satisfying the covering law. Considering 
Hilbert geometries, Hilbert lattices and propositional systems 
together with suitable (`continuous') morphisms, we obtain a triple 
equivalence of the categories $\HilbGeom$, $\HilbLat$ and 
$\PropSys$. And there is a category $\GenHilb$ of generalized 
Hilbert spaces and continuous semilinear maps, with a functor 
$\GenHilb\to\HilbGeom$. Since a Hilbert geometry is a projective 
geometry with extra structure, and a continuous morphism between 
Hilbert geometries is a particular morphism between (underlying) 
projective geometries, there is a faithful functor 
$\HilbGeom\to\ProjGeom$. Similarly there are faithful functors 
$\HilbLat\to\ProjLat$ and $\GenHilb\to\Vec$ too, and the resulting 
(commutative) diagram of categories and functors is sketched in 
figure \ref{0}.
\par
Then we show in section \ref{F} that a Hilbert 
geometry is irreducible (as a projective geometry, i.e.\ each line 
contains at least three points) if and only if it is not a 
non-trivial coproduct in $\HilbGeom$; and each Hilbert geometry is 
the coproduct of irreducible ones. By categorical equivalence, the 
``same'' is true for Hilbert lattices and propositional systems.
\par
In section \ref{G} we present the Representation Theorem for 
propositional systems or, equivalently, Hilbert geometries (of 
dimension at least 2):  
the arguesian Hilbert geometries constitute the image of 
the functor $\GenHilb\to\HilbGeom$. For finite dimensional 
geometries this result is due to G. Birkhoff and J. von Neumann 
[1936] while the more general (infinite-dimensional) version goes 
back to C. Piron's [1964, 1976] representation theorem: every 
irreducible propositional system of rank at least 4 is isomorphic to 
the lattice of closed subspaces of an essentially unique generalized 
Hilbert space. We provide an outline of the proof given in [Holland, 
1995, \S3]. 
\par
The final section \ref{H} contains some comments and remarks on 
various interesting points that we did not address or develop in the 
text.
\\[2mm]
{\bf Required lattice and category theory.} Throughout this chapter 
we use quite a few notions and (mostly straightforward) facts from 
lattice theory. For completeness' sake we have added a short 
appendix in which we explain the words marked with a ``$\dagger$'' 
in our text. The standard references on lattice theory are 
[Birkhoff, 1967; Gr\"atzer, 1998], but [Maeda and Maeda, 1970; 
Kalmbach, 1983] have everything we need too. Finally, we also use 
some very basic category theory: we speak of an `equivalence of 
categories', compute some `coproducts', and talk about `full' and 
`faithful' functors. Other categorical notions that we need, are 
explained in the text. The classic [Mac Lane, 1971] or the first 
volume of [Borceux, 1994] contain all this (and much more).
\\[2mm]
{\bf Acknowledgements.} As students of the '98 generation in 
mathematics in Brussels, both authors prepared a diploma 
dissertation on topics related to operational quantum logic, 
supervised and surrounded by some of the field's most outstanding 
researchers---Dirk Aerts, Bob Coecke, Frank Valckenborgh. Moreover, 
the quantum physics group in Brussels being next of kin to 
Constantin Piron's group in Geneva, we also had the chance to 
interact with the members of the latter---Claude-Alain Faure, 
Constantin Piron, David Moore. It is with great pleasure that we 
dedicate this chapter to all those who made that period 
unforgettable. We thank Mathieu Dupont, Claude-Alain Faure, Chris 
Heunen and Frank Valckenborgh for their comments and suggestions.

\section{Projective geometries, projective lattices}\label{B}
\setcounter{theorem}{0}
It is a well-known slogan in mathematics that ``the lines of a 
vector space are the points of a projective geometry''. To make this 
statement precise, we must introduce the abstract notion of a 
`projective geometry'.
\begin{definition}\label{1}
A {\bf projective geometry}\index{projective geometry} $(G,l)$ is a 
set $G$ of {\bf points} together with a ternary {\bf collinearity 
relation}\index{collinearity relation} $l\subseteq G\times G\times 
G$ such that
\begin{enumerate}
%
%
\setlength{\itemsep}{0pt} \setlength{\parskip}{0pt} 
\item[(G1)] for all $a,b\in G$, $l(a,b,a)$,
\item[(G2)] for all $a,b,p,q\in G$, if $l(a,p,q)$, $l(b,p,q)$ and $p\neq q$, then $l(a,b,p)$,
\item[(G3)] for all $a,b,c,d,p\in G$, if $l(p,a,b)$ and $l(p,c,d)$ then there exists a $q\in G$ such that $l(q,a,c)$ and $l(q,b,d)$.
\end{enumerate}
\end{definition}
Often, since no confusion will arise, we shall speak of ``a projective 
geometry $G$'', without explicitly mentioning its collinearity 
relation $l$. The axioms for the collinearity relation -- as well as 
many of the calculations further on -- are best understood by means 
of a simple picture, in which one draws ``dots'' for the points of 
$G$, and a ``line'' through any three points $a,b,c$ such that 
$l(a,b,c)$. With this intuition (which will be made exact further 
on), (G1) and (G2) say that two distinct points determine one and 
only one line, and (G3) is depicted in figure \ref{1.1}.
\begin{figure}
\begin{center}
\begin{picture}(50,30)
\drawline(0,0)(30,30) \dashline{1}(0,0)(50,0) \drawline(20,0)(30,30) 
\dashline{1}(15,15)(50,0)
\put(0,0){\dot} \put(20,0){\dot} \put(15,15){\dot} \put(50,0){\dot} 
\put(23.7,11.1){\dot} \put(30,30){\dot}
\put(0,3){\lett{a}} \put(19,3){\lett{c}} \put(50,3){\lett{q}} 
\put(15,18){\lett{b}} \put(30,33){\lett{p}} 
\put(22.7,14.1){\lett{d}}
\end{picture}
\end{center}
\caption{Illustration of (G3)} \label{1.1}
\end{figure}
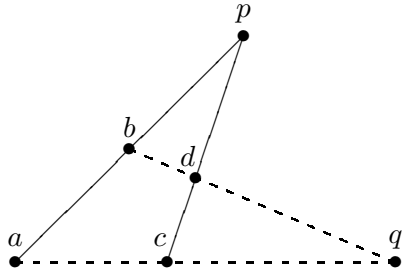
\begin{example}\label{2}
Let $V$ be a (left) vector space over a (not necessarily 
commutative) field $K$. The set of lines of $V$ endowed with the 
coplanarity relation forms a projective geometry\index{projective 
geometry!induced by a vector space}; it will be denoted further on 
as $\P(V)$. Note that the collinearity relation is trivial when 
$\dim(V)\leq 2$.
\end{example}
The example $\P(V)$ is very helpful for sharpening the intuition on 
abstract projective geometry. For example, it is clear that the 
collinearity relation in $\P(V)$ is symmetric; but in fact this 
property holds also in the general case.
\begin{lemma}\label{2.1}
A ternary relation $l$ on a set $G$ satisfying (G1--2) is symmetric, 
meaning that for $a_1,a_2,a_3\in G$, if $l(a_1,a_2,a_3)$ then also 
$l(a_{\sigma(1)},a_{\sigma(2)},a_{\sigma(3)})$ for any permutation 
$\sigma$ on $\{1,2,3\}$.
\end{lemma}   
\begin{proof} The group of permutations on $\{1,2,3\}$ being generated by 
its elements $(1 2 3)\mapsto(132)$ and $(123)\mapsto(312)$, we only 
need to check two cases. This is a simple exercise.
%
%
%
\end{proof}
It is not hard to show that (G3) follows from (G1--2) when 
$\card\{a,b,c,d,p\}\neq 5$ or when $\{a,b,c,d\}$ contains three 
(different) points that belong to $l$.
\par
For a projective geometry $(G,l)$, any two distinct points $a,b\in 
G$ determine the {\bf projective line}\index{projective line} $a\* 
b\defeq\{x\in G\mid l(x,a,b)\}$. For notational convenience, we also 
put that $a\* a\defeq\{a\}$. It is a useful corollary of \ref{2.1} 
that for $a,b,c\in G$, if $a\neq c$ then $a\in b\*c$ implies $b\in 
a\* c$.
%
%
\par
Now we define a {\bf subspace}\index{subspace} $S$ of $G$ to be a 
subset $S\subseteq G$ with the property that
$$\mbox{if }a,b\in S\mbox{ then }a\* b\subseteq S.$$
Trivially, any projective geometry $G$ has the empty subspace 
$\emptyset\subseteq G$ and the total subspace $G\subseteq G$. 
Moreover, all $a\* b\subseteq G$ are subspaces; these include all 
singletons $\{a\}=a\* a$.
\par 
The set of all subspaces of $G$ will be denoted $\L(G)$. Since 
subspaces of $G$ are particular subsets of $G$, $\L(G)$ is ordered 
by inclusion. The following proposition collects some features of 
the ordered set\asterix\ $(\L(G),\subseteq)$, but first we shall 
record a key lemma.
\begin{lemma}\label{6.1}
In the lattice\asterix\ $\L(G)$ of subspaces of a projective 
geometry $G$,
\begin{enumerate}
%
%
\setlength{\itemsep}{0pt} \setlength{\parskip}{0pt} 
\item for any family of subspaces $(S_i)_{i\in I}$, $\bigcap_iS_i$ is a subspace,
\item for a directed\asterix\ family of subspaces $(S_i)_{i\in I}$, $\bigcup_i S_i$ is a subspace,
\item for two non-empty subspaces $S$ and $T$, $\bigcup\{a\* b\mid
  a\in S, b\in T\}$ is a subspace.
\end{enumerate}
\end{lemma}
\begin{proof} The proofs of the first two statements are straightforward. 
As for the third statement, we must prove that, if $l(x,a_1,b_1)$, 
$l(y,a_2,b_2)$ and $l(z,x,y)$, with $a_1,a_2\in S$ and $b_1,b_2\in 
T$, then $l(z,a_3,b_3)$ for some $a_3\in S$ and $b_3\in T$. The 
picture in figure \ref{6.1.1} suggests how to do this, using the 
symmetry of the collinearity relation and applying (G3) over and 
over again.
\begin{figure}
\begin{center}
\begin{picture}(40,40)
\drawline(0,40)(40,40) \drawline(0,20)(40,20) \drawline(0,0)(0,40) 
\drawline(0,0)(40,0) \drawline(20,0)(20,40) \drawline(40,0)(40,40) 
\drawline(20,0)(40,40) \drawline(0,0)(40,20) 
\drawline(0,40)(26.66,13.33)
\put(-3,38){\lett{a_1}} \put(-3,18){\lett{a_2}} 
\put(-2,-2){\lett{a_3}} \put(18,38){\lett{x}} \put(18,18){\lett{y}} 
\put(18,-2){\lett{z}} \put(43,38){\lett{b_1}} 
\put(43,18){\lett{b_2}} \put(43,-2){\lett{b_3}} 
\put(28.66,11.33){\lett{u}}
\put(0,40){\dot} \put(0,20){\dot} \put(0,0){\dot} \put(0,40){\dot} 
\put(20,40){\dot} \put(20,20){\dot} \put(20,0){\dot} 
\put(40,40){\dot} \put(40,20){\dot} \put(40,0){\dot} 
\put(26.66,13.33){\dot}
\end{picture}
\end{center}
\caption{Illustration of the proof of \ref{6.1}} \label{6.1.1}
\end{figure}
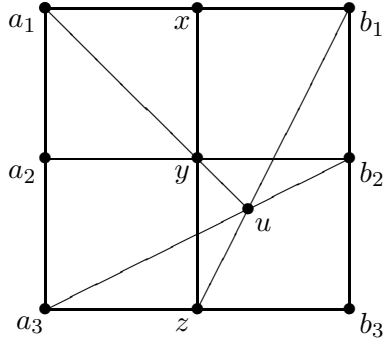
%
%
%
\end{proof}
It follows that for a subset $A\subseteq G$ of a projective geometry 
$G$
$$\cl(A)\defeq\bigcap\{S\in\L(G)\mid A\subseteq S\}$$
is the smallest subspace of $G$ that contains $A$: it is its 
so-called {\bf projective closure}\footnote{This terminology is 
well-chosen, for the mapping $A\mapsto \cl(A)$ does indeed define a 
closure operator\asterix\ on the set of subsets of $G$; see also 
\ref{81}.}\index{projective closure}. The third statement in 
\ref{6.1} is often referred to as the {\bf projective 
law}\index{projective law}. In terms of the projective closure it 
may be stated as: for non-empty subspaces $S$ and $T$ of $G$, 
$$\cl(S\cup T)=\bigcup\{a\* b\mid a\in S, b\in T\}.$$ 
\begin{proposition}\label{6}
For any projective geometry $G$, $(\L(G),\subseteq)$ is a 
complete\asterix, atom\-is\-tic\asterix\index{atomistic lattice}, 
continuous\asterix\index{continuous lattice}, modular\asterix\ 
\index{modular lattice} lattice.\index{projective lattice!induced by 
a projective geometry}
\end{proposition}
\begin{proof} The order on $\L(G)$ is complete, because the intersection of 
subspaces is their infimum\asterix; thus the supremum\asterix\ of a 
family $(S_i)_{i\in I}\in\L(G)$ is $\bigvee_iS_i=\cl(\bigcup_i 
S_i)$. This makes it at once clear that any subspace $S\in\L(G)$ is 
the supremum of its points: $S=\cl(S)=\bigvee_{a\in S}\{a\}$; and 
singleton subspaces being exactly the atoms\asterix\ of $\L(G)$ this 
also shows that $\L(G)$ is atomistic. The continuity of $\L(G)$ 
follows trivially from the fact that directed suprema in $\L(G)$ are 
simply unions. Finally, to show that $\L(G)$ is modular, it suffices 
to verify that for non-empty subspaces $S,T,U\subseteq G$, if 
$S\subseteq T$ then $(S\vee U)\cap T\subseteq S\vee(U\cap T)$. We 
are going to use the projective law a couple of times. Suppose that 
$x\in (S\vee U)\cap T$; so $x\in T$, but also $x\in S\vee U$, which 
means that $x\in a\* b$ for some $a\in S$ and $b\in U$. If $x=a$ 
then $x\in S\subseteq S\vee(U\cap T)$; if $x\neq a$ then $x\in a\* 
b$ implies that $b\in a\* x\subseteq S\vee T=T$ (using that 
$S\subseteq T$) so that $x\in a\* b\subseteq S\vee(U\cap T)$ in this 
case too.
\end{proof}
\begin{definition}\label{7}
An ordered set $(L,\leq)$ is a {\bf projective 
lattice}\index{projective lattice} if it is a complete, atomistic, 
continuous, modular lattice.
\end{definition}
There are equivalent formulations for the definition of `projective 
lattice'; we shall encounter some further on in this section. Here 
we shall already give one alternative for the continuity condition, 
which is sometimes easier to handle and will be used in the proofs 
of \ref{17} and \ref{19.10}.
\begin{lemma}\label{8.1}
A complete atomistic lattice $L$ is continuous\index{continuous 
lattice} if and only if its atoms are {\bf compact}, i.e.\ if $a$ is 
an atom and $(x_i)_{i\in I}$ is a directed family in $L$, then 
$a\leq\bigvee_ix_i$ implies $a\leq x_k$ for some $k\in I$.
\end{lemma}
\begin{proof} If $L$ is continuous and (with notations as in the statement 
of the lemma) $a\leq\bigvee_ix_i$, then 
$a=a\wedge(\bigvee_ix_i)=\bigvee_i(a\wedge x_i)$, so there must be a 
$k\in I$ for which $a\wedge x_k\neq 0$, whence $a\leq x_k$ (for $a$ 
is an atom). Conversely, $\bigvee_i(y\wedge x_i)\leq 
y\wedge(\bigvee_ix_i)$ holds for any element $y\in L$. Suppose that 
this inequality is {\it strict}. By atomisticity of $L$ there must 
exist an atom $a$ such that $a\not\leq\bigvee_i(y\wedge x_i)$ and 
$a\leq y\wedge(\bigvee_ix_i)$. This implies in particular that 
$a\leq y$ and $a\leq\bigvee_ix_i$, and by hypothesis $a$ is compact 
so that $a\leq x_k$ for some $k\in I$. But then $a\leq y\wedge 
x_k\leq\bigvee_i(y\wedge x_i)$ is a contradiction. Thus necessarily 
$\bigvee_i(y\wedge x_i)=y\wedge(\bigvee_ix_i)$ for every $y\in L$.
\end{proof}
\begin{example}\label{8}
For a $K$-vector space $V$, the set $\L(V)$ of linear 
subspaces\index{projective lattice!induced by a vector space}, 
ordered by set-inclusion, is isomorphic to the projective lattice 
$\L(\P(V))$ of subspaces of the projective geometry $\P(V)$ of 
\ref{2}: the mappings
\begin{eqnarray*}
 & & \L(V)\to\L(\P(V))\:W\mapsto\P(W) \\
 & & \L(\P(V))\to\L(V)\:S\mapsto\{x\in V\mid Kx\in S\}\cup\{0\}
\end{eqnarray*}
are well-defined, preserve order\asterix\ and are each other's 
inverse. With slight abuse of notation we shall write $\L(V)$ even 
when we actually mean $\L(\P(V))$.
\end{example}
\par
So \ref{6} states that a projective geometry $G$ determines a 
projective lattice $\L(G)$; but the converse is also true. First we 
prove a lattice theoretical lemma that exhibits the strength of the 
modularity condition.
\begin{lemma}\label{9.1}
Let $L$ be a complete atomistic lattice, and $\G(L)$ its set of 
atoms. 
\begin{enumerate}
%
%
\setlength{\itemsep}{0pt} \setlength{\parskip}{0pt} 
\item\label{i} If $L$ is modular\index{modular lattice}, then it is
  both upper semimodular\asterix\ \index{modular lattice!upper semi-}
  and lower semimodular\asterix\index{modular lattice!lower semi-}.
\item\label{ii} $L$ is upper semimodular if and only if it satisfies the covering law\asterix\index{covering law}.
\item\label{iii} If $L$ is lower semimodular and satisfies the covering law then it has the {\bf intersection property}\index{intersection property}: for any $x\in L$ and $p,q\in\G(L)$ with $p\neq q$, if $p\leq q\vee x$ then $(p\vee q)\wedge x\neq 0$.
\item\label{iv} If $L$ has the intersection property, then for $a,b,c\in\G(L)$ with $a\neq b$, if $a\leq b\vee c$ then also $c\leq a\vee b$.
\item\label{v} If $L$ has the intersection property, then $\G(L)$ forms a projective geometry for the ternary relation\index{collinearity relation}
\begin{equation}\label{9.2}
l(a,b,c)\mbox{ if and only if }a\leq b\vee c\mbox{ or }b=c.
\end{equation}
\end{enumerate}
\end{lemma}
\begin{proof} (\ref{i}) For any lattice $L$, the maps
$$\phi\:[u\wedge v,v]\to[u,u\vee v]\:x\mapsto x\vee u\mbox{ and }
\psi\:[u,u\vee v]\to[u\wedge v,v]\:y\mapsto y\wedge v$$ are 
well-defined and preserve order. If $L$ is modular then moreover 
$\psi(\phi(x))=(x\vee u)\wedge v=x\vee(u\wedge x)=x$; similarly 
$\phi(\psi(y))=y$. So the two segments are isomorphic lattices. Now 
clearly $u\wedge v\cov v\iff \card[u\wedge v,v]=2\iff\card[u,u\vee 
v]=2\iff u\cov u\vee v$, which proves both upper and lower 
semimodularity of $L$ (resp. $\Rightarrow$ and $\Leftarrow$ in this 
equivalence).
\par
(\ref{ii}) The covering law is a special case of upper 
semimodularity. Conversely, in an atomistic lattice satisfying the 
covering law it is the case that 
\begin{equation}\label{9.0.1}
x\cov y\mbox{ if and only if there exists }a\in\G(L): a\not\leq x, 
x\vee a=y.
\end{equation}
(Indeed, the ``only if'' follows from atomisticity, and the ``if'' 
is the covering law.) So if now $u\wedge v\cov v$ then there is an 
atom $a\in\G(L)$ such that $a\not\leq u\wedge v$ and $(u\wedge 
v)\vee a=v$. But then $u\vee v=u\vee[(u\wedge v)\vee 
a)]=[u\vee(u\wedge v)]\vee a=u\vee a$; and $a\not\leq u$ (for $a\leq v$ but $a\not\leq u\wedge v$) so by 
(\ref{9.0.1}) we conclude that $u\cov u\vee v$.
\par
(\ref{iii}) Let $p\neq q\in \G(L)$ and $x\in L$ be such that $p\leq 
q\vee x$. If $q\leq x$ then trivially $q\leq (p\vee q)\wedge x$; if 
$q\not\leq x$ then $x\cov q\vee x=(p\vee q)\vee x$ by the covering 
law and the hypothesis $p\leq q\vee x$. This in turn implies 
$x\wedge(p\vee q)\cov p\vee q$ by lower semimodularity. Now 
$x\wedge(p\vee q)\neq 0$ because it is covered by $p\vee q$ which is 
not an atom.
\par
(\ref{iv}) From the assumptions and the intersection property it 
follows that $(a\vee b)\wedge c\neq 0$, so that necessarily $c\leq 
a\vee b$, for $c$ is an atom. 
\par
(\ref{v}) We shall check the axioms in \ref{1}, using the notations 
introduced there and keeping in mind that the collinearity relation 
is as in (\ref{9.2}). Axiom (G1) is trivial. For (G2) we may suppose 
that $b\neq p$. Then, by (\ref{iv}), $b\leq p\vee q$ implies $q\leq 
b\vee p$ and hence $a\leq p\vee q\leq p\vee (b\vee p)=p\vee b$, as 
wanted. As for (G3), we may suppose that $a$, $b$, $c$, $d$, and $p$ 
are different points; then $p\leq a\vee b$ implies $a\leq p\vee b$, 
hence $a\leq b\vee c\vee d$, and therefore, by the intersection 
property, $(b\vee d)\wedge(a\vee c)\neq 0$, which means (by 
atomisticity) that $q\leq (b\vee d)\wedge(a\vee c)$ for some 
$q\in\G(L)$, as wanted.
\end{proof}
The lemma above is not stated as ``sharply'' as possible. In fact, 
the `intersection property' in (\ref{iii}) can be rephrased for an 
arbitrary lattice $L$ with $0$ as
$$\mbox{if }p\leq q\vee x\mbox{ then there exists an atom $r$ such that }r\leq(p\vee q)\wedge x$$
for atoms $p\neq q$ and an arbitrary $x$; this is how it first 
appeared in [Faure and Fr\"olicher, 1995]. Then statement (\ref{i}), 
sufficiency in (\ref{ii}), and statement (\ref{iii}) are true for 
any lattice with $0$ (not necessarily complete nor atomistic), while 
the converse of (\ref{iii}) holds for an atomistic $L$. We shall 
only need these results for a complete atomistic lattice $L$ (in 
\ref{9} below and also in \ref{108} further on).
\par
We may now state the following as a simple corollary of \ref{9.1}.
\begin{proposition}\label{9}
The set $\G(L)$ of atoms of a projective lattice $L$ forms a 
projective geometry\index{projective geometry!induced by a 
projective lattice} for the ternary relation in (\ref{9.2}).
\end{proposition}
\par
If $L$ is a lattice with $0$ for which $\G(L)$ is a projective
geometry for the collinearity in (\ref{9.2}), then $\L(\G(L))$ is a
projective lattice, according to \ref{6}. Would $L$ be isomorphic to
$\L(\G(L))$ then necessarily $L$ must be a projective lattice too:
completeness, atomisticity, continuity and modularity are transported
by isomorphism. From the work in the rest of this section it will
follow that $L$ being projective is also sufficient for it to be
naturally isomorphic to $\L(\G(L))$. Similarly it is also true that a
projective geometry $G$ may be identified with $\G(\L(G))$. More
precisely, we shall show that projective geometries and projective
lattices are categorically equivalent notions. So we better start
building categories!
\par
Recall first that a {\bf partial map} $f$ between sets $A$ and $B$ 
is a map from a subset $D_f\subseteq A$ to $B$. The set $D_f$ is the 
{\bf domain} of $f$, and the set-complement $K_f=(D_f)\c$ is its 
{\bf kernel}. Most of the time we write such a partial map as 
$f\:A\parto B$ instead of $f\:A\setminus K_f\to B$ or 
$f\:D_f\subseteq A\to B$. Partial maps compose: for $f\:A\parto B$ 
with kernel $K_f$ and $g\:B\parto C$ with kernel $K_g$, $g\circ 
f\:A\parto C$ has kernel $K_f\cup f^{-1}(K_g)$ and maps an element 
$a$ of its domain to $g(f(a))$. This composition law is associative, 
and the identity map on a set (viewed as partial map with empty 
kernel) is a two-sided identity for this composition. That is to 
say, there is a perfectly good category $\ParSet$ of sets and 
partial maps.
\begin{definition}\label{11}
Given two projective geometries $G_1$ and $G_2$, a partial map 
$g\:G_1\parto G_2$ is a {\bf morphism of projective 
geometries}\index{morphism!of projective geometries} if, for any 
subspace $T$ of $G_2$,
$$g^*(T)\defeq K_g\cup g^{-1}(T)$$
is a subspace of $G_1$.
\end{definition}
Since $\emptyset\subseteq G_2$ is a subspace, the kernel $K_g$ of 
$g\:G_1\parto G_2$ must be a subspace of $G_1$. In the proof of 
\ref{19.9.1} we shall show that a morphism $g\:G_1\parto G_2$ maps 
any line $a\* b$ in $G_1$, with $a,b\not\in K_g$, either to a single 
point of $G_2$ (in case $g(a)=g(b)$) or injectively to the line 
$g(a)\* g(b)$ of $G_2$ (in case $g(a)\neq g(b)$). This provides a 
geometric interpretation, in terms of points and lines, of the 
notion of `morphism between projective geometries'. (As a matter of 
fact, these latter conditions are also sufficient for $g$ to be a 
morphism, provided that $K_g=\emptyset$.) 
\par
With composition of two morphisms of projective geometries defined 
as the composition of the underlying partial maps, we obtain a 
category\index{projective geometry!category of -ies} $\ProjGeom$. An 
isomorphism in $\ProjGeom$ is, as in any category, a morphism 
$g\:G_1\parto G_2$ with a two-sided inverse $g'\:G_2\parto G_1$. But 
it can easily be seen that such is the same as a bijection (with 
empty kernel) $g\:G_1\to G_2$ which preserves and reflects the 
collinearity relation: $l_1(a,b,c)$ if and only if 
$l_2(g(a),g(b),g(c))$ for all $a,b,c\in G_1$.
\begin{example}\label{12}
By definition, a \textbf{semilinear map}\index{semilinear map} 
between a $K_1$-vector space $V_1$ and a $K_2$-vector space $V_2$ is 
an additive map $f\:V_1\to V_2$ for which there exists a 
homomorphism of fields $\sigma\:K_1\to K_2$ such that $f(\alpha 
x)=\sigma(\alpha)f(x)$ for every $\alpha\in K_1$ and $x\in V_1$. 
Sometimes we call this a \textbf{$\sigma$-linear 
map}\index{$\sigma$-linear map|see{semilinear map}} $f\:V_1\to V_2$ 
too. The $\sigma$ is uniquely determined by $f$ whenever $f$ is 
non-zero; and the zero-map is semilinear if and only if there exists 
a homomorphism $\sigma\:K_1\to K_2$. There is a 
category\index{vector space!category of -s} $\SemiVec$ of vector 
spaces and semilinear maps. A semilinear map $f\:V_1\to V_2$ 
determines a morphism of projective geometries\index{morphism!of 
projective geometries}
$$\P(f)\:\P(V_1)\parto\P(V_2)\:K_1x\mapsto K_2f(x)\mbox{ with kernel }\P(\ker(f)).$$
This, in fact, defines a functor 
$$\P\:\SemiVec\to\ProjGeom\:\Big(f\:V_1\to V_2\Big)\mapsto\Big(\P(f)\:\P(V_1)\parto\P(V_2)\Big).$$
\end{example}
The following example [Faure and Fr\"olicher, 2000, 6.3.9--11] shows 
that semilinear maps can behave surprisingly when the associated 
field homomorphism is not an isomorphism.
\begin{example}\label{12.1}
Let $F$ be a commutative field, let $K\defeq F(x)$ be the field of 
rational functions and let $n$ be a positive integer. Consider the 
field homomorphism $\sigma\: K\to K\: q(x) \mapsto q(x^n)$. One can 
show that $\sigma(K) \subseteq K$ is an extension of fields of 
degree $n$: putting $\alpha_i \defeq x^{i-1}$, the set 
$\{\alpha_1,\dots,\alpha_n\}$ forms a basis of $K$ over $\sigma(K)$. 
It follows that $\varphi\: K^n\to K\: 
(a_1,\dots,a_n)\mapsto\sigma(a_1)\alpha_1+\dots+\sigma(a_n)\alpha_n$ 
is a $\sigma$-linear form with zero kernel. Moreover, picking any 
nonzero $b \in K^n$ we obtain a $\sigma$-linear map $f\: K^n\to 
K^n\: x\mapsto\varphi(x)b$ for which $\P(f)$ is constant and which has 
empty kernel.
\end{example}
\par
We now turn to projective lattices.
\begin{definition}\label{13}
Given projective lattices $L_1$ and $L_2$, a map $f\:L_1\to L_2$ is 
a {\bf morphism of projective lattices}\index{morphism!of projective 
lattices} if it preserves arbitrary suprema and sends atoms in $L_1$ 
to atoms or to the bottom element in $L_2$.
\end{definition}
We thus get a category\index{projective lattice!category of -s} 
$\ProjLat$. Note that an isomorphism in $\ProjLat$ is indeed the 
same thing as an order-preserving and reflecting bijection, so that 
in \ref{8} there is no doubt about the meaning of the word.
\par
We know from \ref{6} that any projective geometry $G$ determines a 
projective lattice $\L(G)$; and any projective lattice $L$ 
determines a projective geometry $\G(L)$ according to \ref{9}. For 
morphisms we can play a similar game.\index{morphism!of projective 
geometries}\index{morphism!of projective lattices}
\begin{proposition}\label{14}
If $g\:G_1\parto G_2$ is a morphism of projective geometries, then 
$$\L(g)\:\L(G_1)\to\L(G_2)\:S\mapsto \bigcap\{T\in\L(G_2)\mid S\subseteq g^*(T)\}$$
is a morphism of projective lattices. And if $f\:L_1\to L_2$ is a 
morphism of projective lattices, then
$$\G(f)\:\G(L_1)\parto\G(L_2)\:a\mapsto f(a)\mbox{ with kernel }\{a\in\G(L_1)\mid f(a)=0\}$$
is a morphism of projective geometries.
\end{proposition}
\begin{proof} First note that $g\:G_1\parto G_2$ defines the ``inverse image'' map
$$g^*\:\L(G_2)\to\L(G_1)\:T\mapsto K_g\cup g^{-1}(T),$$
which preserves arbitrary intersections. Intersections of 
subspaces being their infima, $g^*$ must have a left adjoint\asterix. This 
left adjoint is precisely $\L(g)$, which proves that $\L(g)$ 
preserves arbitrary suprema. The atoms of $\L(G_1)$ and $\L(G_2)$ 
corresponding to their respective singleton subspaces, $\L(g)$ sends 
atoms to atoms or to the bottom element.
\par
Because $f\:L_1\to L_2$ sends atoms of $L_1$ to atoms or the bottom 
element of $L_2$, $\G(f)$ is a well-defined partial map. Now let 
$T\subseteq\G(L_2)$ be a subspace of the projective geometry 
$\G(L_2)$; if $a, b\in\G(f)^*(T)$ and $c\leq a\vee b$ then $f(c)\leq 
f(a\vee b)=f(a)\vee f(b)$, showing that either $f(c)=0$ or $f(c)\in 
T$ (by $T$ being a subspace). That is to say, $\G(f)^*(T)$ is a 
subspace of $G_1$.
\end{proof}
The above proposition explains the requirement in \ref{13} that a 
morphism of projective lattices preserve {\it arbitrary} suprema: 
such a morphism must be thought of as the left adjoint to an inverse 
image.
\par
Now we are ready to state and prove the result promised a while ago.
\begin{theorem}\label{17}
The categories $\ProjGeom$ and $\ProjLat$ are 
equivalent\index{equivalence of categories!of projective geometries 
and projective lattices}. To wit, the assignments
\begin{eqnarray*}
 & & \L\:\ProjGeom\to\ProjLat\:\Big(g\:G_1\parto G_2\Big)\mapsto\Big(\L(g)\:\L(G_1)\to \L(G_2)\Big) \\
 & & \G\:\ProjLat\to\ProjGeom\:\Big(f\:L_1\to L_2\Big)\mapsto\Big(\G(f)\:\G(L_1)\parto\G(L_2)\Big)
\end{eqnarray*}
are functorial, and for a projective geometry $G$ and a projective 
lattice $L$ there are natural isomorphisms
\begin{eqnarray*}
 & & \alpha_G\:G\iso\G(\L(G))\:a\mapsto\{a\}, \\
 & & \beta_L\:L\iso\L(\G(L))\:x\mapsto\{a\in\G(L)\mid a\leq x\}.
\end{eqnarray*}
\end{theorem}
\begin{proof} It is a matter of straightforward calculations to see that 
$\L$ and $\G$ are functorial. We shall prove that $\alpha_G$ and 
$\beta_L$ are isomorphisms, and leave the verification of their 
naturality to the reader.
\par
First, the map $\alpha_G$ is obviously a well-defined bijection 
(with empty kernel): the atoms of $\L(G)$ are precisely the 
singleton subsets of $G$, i.e.\ the points of $G$. We need to show 
that $a,b,c$ are collinear in $G$ if and only if $\{a\},\{b\},\{c\}$ 
are collinear in $\G(\L(G))$; but this comes down to showing that 
$a\in b\*c$ in $G$ if and only if $\{a\}\subseteq \{b\}\vee \{c\}$ 
in $\L(G)$, which is an instance of the projective law.
\par
Next, it is easy to see that $\beta_L$ is a well-defined map, i.e.\ 
that any $\beta_L(x)\subseteq \G(L)$ is indeed a subspace (for the 
collinearity relation on $\G(L)$ as in \ref{9}). We claim now that 
the map
$$\gamma_L\:\L(\G(L))\mapsto L\:S\mapsto\bigvee S$$
is the inverse of $\beta_L$ in $\ProjLat$. In fact, it is clear that 
both $\beta_L$ and $\gamma_L$ preserve order; thus it suffices to 
show that they are mutually inverse maps to prove that they 
constitute an isomorphism in $\ProjLat$. That $\gamma_L\circ\beta_L$ 
is the identity, is the atomisticity of $L$. Conversely, for a 
subspace $S\subseteq\G(L)$ we have that 
$(\beta_L\circ\gamma_L)(S)=\{a\in\G(L)\mid a\leq\bigvee S\}$; so it 
suffices to prove that $a\leq\bigvee S\iff a\in S$ to find that 
$\beta_L\circ\gamma_L$ is the identity on $\L(\G(L))$. But $\bigvee 
S=\bigvee\{\bigvee S'\mid S'\subseteq\f S\}$ -- where we write 
$S'\subseteq\f S$ for a {\it finite} subset -- which expresses 
$\bigvee S$ as a directed join of finite joins. Because the atoms of 
$L$ are compact (by continuity and \ref{8.1}), $a\leq\bigvee S$ if 
and only if $a\leq\bigvee S'$ for some $S'\subseteq\f S$. Using the 
intersection property of $L$ (cf.\ \ref{9.1} and \ref{9}) and using 
the subspace property of $S$ for the collinearity relation on 
$\G(L)$, we shall prove by induction on the number of elements of 
$S'=\{s_1,...,s_n\}$ that $a\in S$. The case $n=1$ is trivial, so 
let the case $n-1$ be true by induction hypothesis, and let $a\leq 
s_1\vee ...\vee s_{n-1}\vee s_n$ with $s_n\not\leq s_1\vee...\vee 
s_{n-1}$. If $a\leq s_n$ then $a=s_n\in S$ and we are done. If 
$a\not\leq s_n$ then $a\neq s_n$ and by the intersection property 
$(s_1\vee...\vee s_{n-1})\wedge(s_n\vee a)\neq 0$, so (by 
atomisticity) there is an atom $r\in\G(L)$ such that 
$r\leq(s_1\vee...\vee s_{n-1})\wedge(s_n\vee a)$. But then $r\leq 
s_1\vee...\vee s_{n-1}$ thus $r\in S$ by the induction hypothesis. 
And since $r\neq s_n$ (for otherwise $s_n\leq s_1\vee...\vee 
s_{n-1}$), $r\leq s_n\vee a$ implies $a\leq r\vee s_n$ by (\ref{iv}) 
of \ref{9.1}, so that $a\in S$ by $S$ being a subspace.
\end{proof}
In the proof for $\beta_L\:L\to\L(\G(L))$ being an isomorphism, 
modularity of $L$ was not explicitly used, except for the fact that 
it implies the intersection property as in \ref{9.1}. Therefore, 
since $L\cong\L(\G(L))$ and the latter is a projective lattice, it 
follows that a complete, atomistic, continuous lattice is modular if 
and only if it has the intersection property.
\par
Until now we have considered the following diagram of categories and 
functors:
$$\xymatrix{\Vec\ar[r]& \ProjGeom\ar@{<->}[r]^{\ \ \sim}& \ProjLat.}$$
In section \ref{D} we shall discuss a converse to the functor 
$\P\:\SemiVec\to\ProjGeom$, but thereto we need to deal with another 
issue first.

\section{Irreducible components}\label{C}
\setcounter{theorem}{0}
However trivial it may seem that every plane in a vector space $V$ 
contains at least three lines, this is actually not automatic for 
abstract projective geometries.
\begin{definition}\label{19.1}
A projective geometry $(G,l)$ is {\bf irreducible}\index{projective 
geometry!irreducible} if for every $a,b\in G$, $\card(a\* b)\neq 2$; 
otherwise it is {\bf reducible}.
\end{definition}
Since we defined that $a\* a=\{a\}$, this definition says that $G$ 
is an irreducible projective geometry precisely when every line 
contains at least three points. This definition is clearly invariant 
under isomorphism.
\begin{example}\label{19.2}
For any vector space $V$, $\P(V)$ is an irreducible projective 
geometry: if $K x\neq K y$ then $K(x+y)$ is a third point on the 
line $K x\*K y$. Taking $V$ to be the cube of the field with two 
elements, one gets the smallest irreducible projective geometry with 
three non-collinear points; it is pictured in figure \ref{333} (all 
straight segments {\it and } the circle in the picture designate 
projective lines).
\begin{figure}
\begin{center}
\begin{picture}(30,20)(-15,-5)
\put(0,0){\circle{12.6}} 
\drawline(-11.0851252,-6.4)(11.0851252,-6.4)(0,12.6)(-11.0851252,-6.4) 
\drawline(-11.0851252,-6.4)(5.54,3.2) 
\drawline(11.0851252,-6.4)(-5.54,3.2) \drawline(0,12.6)(0,-6.4)
\put(0,0){\dot} \put(0,12.6){\dot} \put(0,-6.4){\dot} 
\put(-11.0851252,-6.4){\dot} \put(11.0851252,-6.4){\dot} 
\put(5.54,3.2){\dot} \put(-5.54,3.2){\dot}
\end{picture}
\caption{The smallest non-trivial irreducible geometry} \label{333}
\end{center}
\end{figure}
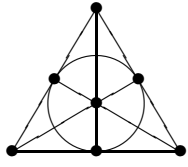
\end{example}
\begin{example}\label{19.3}
Any set $G$ becomes a {\bf discrete projective 
geometry}\index{projective geometry!discrete} when putting 
$l(a,b,c)$ to mean that $\card\{a,b,c\}\leq 2$. A discrete 
projective geometry is irreducible if and only if $G$ is a 
singleton.
\end{example}
The following construction is in a precise sense a generalization of 
\ref{19.3}; it is important enough to record it as a lemma saying in 
particular that the category $\ProjGeom$ has coproducts.
\begin{lemma}\label{19.5}
Given a family 
$(G_i,l_i)_{i\in I}$ of projective geometries, the disjoint union 
$\biguplus_i G_i$ equipped with the relation 
$$l(a,b,c)\mbox{ if either }\card\{a,b,c\}\leq 2\mbox{ or }l_k(a,b,c)\mbox{ in $G_k$ for some }k\in I,$$
together with the inclusions
\begin{equation}\label{19.7}
s_k\:G_k\to\biguplus_iG_i\:a\mapsto a
\end{equation}
is a coproduct in $\ProjGeom$.
\end{lemma}
\begin{proof} First we check that $\biguplus_iG_i$ forms a projective 
geometry for the indicated collinearity relation; we shall verify 
axioms (G1--3) in \ref{1}, keeping the notations used there. For 
(G1) there is nothing to prove. Axiom (G2) is trivial when we have 
$\card\{a,b,p\}\neq 3$; from now on we assume the contrary. If $a=q$ 
then $l(b,p,q)$ means that $b,p,a\in G_k$ and $l_k(b,p,a)$ for some 
$k\in I$ because of the previous assumption, but then $l_k(a,b,p)$ 
by symmetry of $l_k$ and hence $l(a,b,p)$ as wanted; so suppose 
$a\neq q$. The hypothesis $l(a,p,q)$ now implies that $a,p,q\in G_k$ 
and $l_k(a,p,q)$ for some $k\in I$; but then also $b\in G_k$ and 
$l_k(b,p,q)$ by $l(b,p,q)$; so (G2) for $(\biguplus_iG_i,l)$ follows 
from (G2) for $(G_k,l_k)$. Finally, it suffices to check (G3) in the 
case where $\card\{a,b,c,d,p\}=5$; but by the hypotheses $l(p,a,b)$ 
and $l(p,c,d)$ these points must then all lie in the same $G_k$ and 
satisfy $l_k(p,a,b)$ and $l_k(p,c,d)$; so applying (G3) to 
$(G_k,l_k)$ proves (G3) for $(\biguplus_iG_i,l)$.
\par
From the definition of the projective geometry $(\biguplus_iG_i,l)$ 
it follows directly that, for $a,b\in\biguplus_iG_i$, $a\* b=a\*_k 
b$ if $a,b\in G_k$ and $a\* b=\{a,b\}$ otherwise. From this it 
follows in turn that a sub{\it set} $S\subseteq\biguplus_iG_i$ is a 
sub{\it space} if and only if, for every $k\in I$, $S\cap G_k$ is a 
subspace of $G_k$. But then, referring to the maps in (\ref{19.7}), 
since $s_k^*(S)=S\cap G_k$ these maps are morphisms (with empty 
kernels) of projective geometries, forming a cocone in $\ProjGeom$.
\par
Suppose finally that $(g_k\:G_k\parto G)_{k\in I}$ is another cocone 
in $\ProjGeom$; we claim that 
$$g\:\biguplus_iG_i\parto G\:a\mapsto g_k(a)\mbox{ if }a\in G_k\setminus K_{g_k}\mbox{, with kernel }\biguplus_iK_{g_i}$$
is the unique morphism of projective geometries satisfying $g\circ 
s_k=g_k$ for all $k\in I$. To see this, note first that 
$g^*(S)=\biguplus_ig^*_i(S)$ for a subspace $S\subseteq G$; so 
$g^*(S)\cap G_k=g^*_k(S)$ for $k\in I$, and since these are 
subspaces of the respective $G_k$'s, it follows that $g^*(S)$ is a 
subspace of $\biguplus_iG_i$. Hence $g$ is a morphism of projective 
geometries; and obviously $g\circ s_k=g_k$ for all $k\in I$. If 
$\overline{g}\:\biguplus_iG_i\parto G$ is another such morphism, 
then necessarily 
$K_{\overline{g}}=\overline{g}^*(\emptyset)=\biguplus_iK_{g_i}=K_g$; 
and for $a\in G_k\setminus K_{g_k}$, 
$\overline{g}(a)=\overline{g}(s_k(a))=g_k(a)=g(s_k(a))=g(a)$. That 
is to say, $\overline{g}=g$, and we thus verified the universal 
property of the cocone in (\ref{19.7}).
\end{proof}
Clearly, any coproduct of two or more (non-empty) projective 
geometries is reducible. In fact, a discrete projective geometry $G$ 
as in \ref{19.3} is nothing but the coproduct of the singleton 
projective geometries $(\{a\})_{a\in G}$. 
\par
As the terminology suggests, every projective geometry $G$ can be 
``reduced'' to a coproduct of irreducible ones. Note first that a 
subspace $S\subseteq G$ of a projective geometry $(G,l)$ is a 
projective geometry for the inherited collinearity relation; and the 
inclusion $S\hookrightarrow G$ is then a morphism (with empty 
kernel) of projective geometries. We say that $S\subseteq G$ is an 
{\bf irreducible subspace} when it is irreducible as projective 
geometry in its own right; and $S$ is a {\bf maximal irreducible 
subspace} if moreover it is not strictly contained in any other 
irreducible subspace. This terminology is consistent: a projective 
geometry $G$ is irreducible if and only if $G\subseteq G$ is a 
(trivially maximal) irreducible subspace.
\begin{theorem}\label{19.8}
Any projective geometry $G$ is the coproduct\index{subspace!maximal 
irreducible}\index{projective geometry!coproduct decomposition of} 
in $\ProjGeom$ of its maximal irreducible subspaces, which are 
precisely the equivalence classes of the equivalence relation on $G$ 
defined as: $a\sim b$ if $\card(a\* b)\neq 2$.
\end{theorem}
\begin{proof} If $G$ has no points, then there is nothing to prove, so in 
the rest of this proof we suppose that $G\neq\emptyset$.
\par
First we check that the binary relation $\sim$ on $G$ is an 
equivalence relation. Reflexivity and symmetry are trivial; for the 
transitivity we argue as follows (see also figure \ref{19.9.1.1}). 
Suppose that $a,b,c\in G$ are different non-collinear points, that 
$x$ is a third point on the line $a\* b$ and that $y$ is a third 
point on $b\* c$; then there exists a third point $z$ on $a\* c$. 
Indeed, we have from $l(b,a,x)$ and $l(b,c,y)$ that $l(z,a,c)$ and 
$l(z,x,y)$ for some $z\in G$. Would $z=a$ then $l(y,a,x)$, 
$l(b,a,x)$ and $l(y,b,c)$ imply that $y\in(a\* 
x)\cap(b\*c)=(a\*b)\cap(b\* c)=\{b\}$, which is in contradiction 
with the hypothesis that $y\neq b$; so $z\neq a$, and similarly 
$z\neq c$.
\begin{figure}
\begin{center}
\begin{picture}(50,30)
\drawline(0,0)(30,30) \dashline{1}(0,0)(50,15) 
\drawline(30,9)(30,30) \dashline{1}(15,15)(50,15)
\put(0,0){\dot} \put(30,9){\dot} \put(15,15){\dot} \put(50,15){\dot} 
\put(30,15){\dot} \put(30,30){\dot}
\put(0,3){\lett{a}} \put(32,7){\lett{c}} \put(50,18){\lett{z}} 
\put(15,18){\lett{x}} \put(30,33){\lett{b}} \put(32,17){\lett{y}}
\end{picture}
\end{center}
\caption{Illustration for the proof of \ref{19.8}} \label{19.9.1.1}
\end{figure}
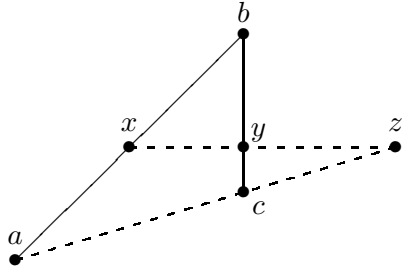
\par
An equivalence class $[a]$ for this relation is a subspace of $G$: if
$x\neq y\in[a]$ and $z\in x\*y$, then either $z=x\in[a]$, or
$z=y\in[a]$, or $\card\{x,y,z\}=3$ and $l(x,y,z)$ so $z\sim x$ thus
$z\in[a]$ by transitivity. In fact, $[a]$ is an irreducible subspace,
because $x,y\in[a]$ implies that $x\sim y$ so that, when $x\neq y$,
automatically $\card(x\* y)\neq 2$. Would $[a]$ be contained in
another irreducible subspace $S\subseteq G$, then $a\* x\subseteq S$
for every $x\in S$ but at the same time $\card(a\* x)\neq 2$; so in
fact $x\sim a$, whence $x\in [a]$. This means that $[a]$ is a maximal
irreducible subspace. Further, if $S\subseteq G$ is {\it any}
non-empty irreducible subspace, then all of its elements are
equivalent, hence $S$ is contained in one of the equivalence classes.
This proves that the latter are {\it precisely} all the maximal
irreducible subspaces of $G$.
\par
Finally, as for any equivalence relation, the equivalence classes of
$\sim$ form a covering by disjoint subsets of $G$. As any subspace,
$[a]$ becomes a projective geometry in its own right for the inherited
collinearity and the inclusions $[a]\hookrightarrow G$ are
morphisms (with empty kernels) of projective geometries.  Moreover, it
is straightforward that, for $x,y,z\in G$, $l(x,y,z)$ if and only if
{\it either} $x,y,z$ are collinear in $[x]$ {\it or}
$\card\{x,y,z\}\leq 2$. By \ref{19.5}, $G$ is thus the coproduct of
the equivalence classes of $\sim$.
\end{proof}
The following categorical characterization of irreducible projective 
geometries is now an easy corollary.
\begin{corollary}\label{19.7.1}
A projective geometry $G$ is irreducible\index{projective 
geometry!irreducible} if and only if it is not a coproduct in 
$\ProjGeom$ of two or more non-empty projective geometries.
\end{corollary}
\par
Interestingly, morphisms of projective geometries behave well with 
respect to irreducible components, as the next proposition shows.
\begin{proposition}\label{19.9.1}
If a morphism of projective geometries $g\:G_1\parto G_2$ has an 
irreducible domain, then its image lies in a maximal irreducible 
subspace of $G_2$. 
\end{proposition}
\begin{proof} If $g(a)\neq g(b)$ for $a,b\in G_1\setminus K_g$, then $a\neq 
b$ so there exists a $c\in G_1$, different from $a$ and $b$, such 
that $c\in a\* b$. We shall show that $g(c)$ is different from 
$g(a)$ and $g(b)$ and lies on $g(a)\* g(b)$, so that $g(a)$ and 
$g(b)$ indeed lie in the same maximal irreducible subspace of $G_2$.
\par
Suppose first that $c\in K_g$; then $b,c\in g^*(\{g(b)\})$ thus 
$a\in b\* c\subseteq g^*(\{g(b)\})$: this is in contradiction with 
$a\not\in K_g$ and $g(a)\neq g(b)$. Hence we know that $c\not\in 
K_g$, and therefore $c\in a\* b\subseteq g^*(g(a)\* g(b))$ implies 
that $g(c)\in g(a)\* g(b)$. Would now $g(c)=g(a)$, then $a,c\in 
g^*(\{g(a)\})$ and this implies a contradiction in the same way as 
before; so $g(c)\neq g(a)$. Similarly one shows that $g(c)\neq 
g(b)$.
\end{proof}
A morphism $g\:G_1\parto G_2$ of projective geometries is, by the 
universal property of the coproduct, the same thing as a family 
$(g^i\:G_1^i\parto G_2)_{i\in I}$ of morphisms, where $(G^i_1)_{i\in 
I}$ denotes the family of maximal irreducible subspaces of $G_1$. 
Writing $(G^j_2)_{j\in J}$ for the family of maximal irreducible 
subspaces of $G_2$, we know by \ref{19.9.1} that each image 
$g^i(G_1^i)$ lies in some $G_2^{j_i}$. Hence $g$ can be ``reduced'' 
to a family $(g^{i}\:G_1^i\parto G_2^{j_i})_{i\in I}$ of morphisms 
between irreducible projective geometries. This goes to show that, 
when studying projective geometry, we can limit our attention to 
irreducible geometries and morphisms between them; after all, the 
reducible ones can be ``regenerated by taking coproducts''.
\par
Since the categories $\ProjGeom$ and $\ProjLat$ are equivalent, the 
previous results on projective geometries have twin siblings for 
projective lattices. We shall go through the translation from 
geometries to lattices. First a word on the construction of 
coproducts of projective lattices.
\begin{lemma}\label{19.10}
For a family of 
projective lattices $(L_i)_{i\in I}$, the cartesian product of sets 
$\times_i L_i$ equipped with componentwise order, together with the 
inclusion maps
\begin{equation}\label{19.11}
s_k\:L_k\to \times_i L_i\:x\mapsto (x_i)_{i\in I}
\end{equation}
where $x_k=x$ and $x_i=0$ for $i\neq k$, is a coproduct in $\ProjLat$.
\end{lemma}
\begin{proof} By the categorical equivalence $\ProjGeom\simeq\ProjLat$ and 
\ref{19.5}, we already know that coproducts exist in $\ProjLat$; we 
shall quickly verify their explicit construction as given in the 
statement of the lemma.
\par
First we check that the cartesian product $\times_iL_i$ is a 
projective lattice whenever the $L_i$'s are\footnote{The converse is 
also true; see \ref{85.1}.}. Since $\times_iL_i$ has the 
componentwise structure, it is clear that it is complete and 
modular; in particular is the zero tuple $0=(0_i)_i$ its least 
element. An atom in $\times_iL_i$ is precisely an element 
$a=(a_i)_i$ with all components zero except for one $a_k$ which is 
an atom in $L_k$; thus it follows easily that $\times_iL_i$ is 
atomistic too. As for the continuity of $\times_iL_i$, it now 
suffices by \ref{8.1} to show that its atoms are compact: but if 
$a\leq\bigvee_{\alpha}x^{\alpha}$ for some atom $a$ and a directed 
family $(x^{\alpha})_{\alpha\in A}$ in $\times_iL_i$, then 
(supposing that the non-zero component of $a=(a_i)_i$ is the atom 
$a_k\in L_k$) necessarily $a_k\leq\bigvee_{\alpha}x^{\alpha}_k$ in 
$L_k$. Since $a_k$ is compact in $L_k$, we have $a_k\leq 
x^{\beta}_k$ for some $\beta\in A$, and thus also $a\leq x^{\beta}$ 
because the components of $a$ other than $a_k$ are zero.
\par
It is a consequence of these observations that the maps in 
(\ref{19.11}) preserve suprema and send atoms onto atoms; thus they 
indeed constitute a cocone in $\ProjLat$. This cocone is universal, 
for if $(f_k\:L_k\to L)_{k\in I}$ is another cocone in $\ProjLat$, 
then the map
$$f\:\times_iL_i\to L\:(x_i)_i\mapsto\bigvee_if_i(x_i)$$
is clearly the unique morphism of projective lattices satisfying 
$f\circ s_k=f_k$ for all $k\in I$.
\end{proof}
Since \ref{19.7.1} tells us ``in categorical terms'' what the 
irreducibility of a projective geometry is all about, the following 
is entirely natural (given that under a categorical equivalence 
coproducts in one category correspond to coproducts in the other).
\begin{definition}\label{x1}
A projective lattice $L$ is {\bf irreducible}\index{projective 
lattice!irreducible} if it is not a coproduct in $\ProjLat$ of two 
(or more) non-trivial projective lattices.
\end{definition}
\begin{proposition}\label{x1.1}
Let $G$ be a projective geometry and $L$ a projective lattice that 
correspond to each other under the categorical equivalence 
$\ProjGeom\simeq\ProjLat$. Then $G$ is irreducible if and only if 
$L$ is irreducible.
\end{proposition}
One can now deduce, again from the equivalence of projective 
geometries and projective lattices, the following statement.
\begin{theorem}\label{x2}
Each projective lattice $L$ can be written as a 
coproduct\index{projective lattice!coproduct decomposition of} in 
$\ProjLat$ of irreducible projective lattices.
\end{theorem}
We could have given a much more precise statement of the previous 
theorem: it would speak of ``maximal irreducible segments'' of a 
projective lattice as analogs for the maximal irreducible subspaces 
of a projective geometry, and so forth. But we do not really need 
this precision and detail further on, so we shall leave it to the 
interested reader to figure out the {\it exact} analog of 
\ref{19.8}. 
\par
On the other hand, in references on lattice theory such as G. 
Birkhoff's [1967] or F. Maeda and S. Maeda's [1970], the previous 
theorem is often given for a vastly larger class of lattices. 
Thereto one typically makes use of the very general notion of 
`central element' of a (bounded) lattice. This highly interesting 
subject falls outside the scope of this chapter (but see also 
\ref{85.1}). 

\section{The Fundamental Theorems of projective geometry}\label{D}
\setcounter{theorem}{0}
In this section we will explain to what extent the functor 
$\P\:\Vec\to\ProjGeom$ can be ``inverted'': we will describe linear 
representations of projective geometries and the morphisms between 
them. It is a very nice result that the objects and morphisms in the 
image of $\P$ can indeed be characterized geometrically; this is the 
content of the age-old First Fundamental Theorem of projective 
geometry (for the objects) and the more recent\footnote{Calling the 
Second Fundamental Theorem ``recent'' for the case of isomorphisms 
would be quite a stretch: it can be found in [Baer, 1952, chapter 
III, \S1] or [Artin, 1957, chapter II, \S10] for example. However, 
the more general case we present here is due to due to Cl.-A.~Faure 
and A.~Fr\"olicher [1994].} Second Fundamental Theorem (for the 
morphisms). Moreover, it turns out that $\P$ is ``injective up to 
scalar'' on so-called `non-degenerate' semilinear maps (as stated 
explicitly in \ref{27}) and ``injective up to isomorphism'' on 
vector spaces of dimension at least 3 (as in \ref{abc}). 
\par
We will only provide a brief sketch of the proof of the First 
Fundamental Theorem: we essentially outline the proof of R. Baer 
[1952, chapter VII] following the pleasant [Beutelspacher and 
Rosenbaum, 1998, chapter 3] (see also [Maeda and Maeda, 1970, \S 
33--34]).  For the morphisms we refer to the very short [Faure, 
2002], which is inspired by and generalizes [Baer, 1952, \S III.1]. All the 
details can be found in these references or in [Faure and 
Fr\"olicher, 2000, chapters 8, 9 and 10].
\par
In order to state the Fundamental Theorems of projective geometry 
properly, we need to introduce the `dimension' of a projective 
geometry. We refer to [Baer, 1952, VII.2] and [Faure and 
Fr\"olicher, 2000, chapter 4] for more on this. 
\begin{definition}\label{20}
A projective lattice $L$ is {\bf of finite rank}\index{projective 
lattice!rank of} if its top element $1\in L$ is a supremum of a 
finite number of atoms; the {\bf rank} of $L$ is then the minimum 
number of atoms required to write $1$ as their supremum, written as 
$\rk(L)$. Otherwise $L$ is {\bf of infinite rank}, written 
$\rk(L)=\infty$. The {\bf dimension}\index{projective 
geometry!dimension of} of a projective geometry $G$ is 
$\dim(G)\defeq\rk(\L(G))-1$ (which can be $\infty$).
\end{definition}
\begin{example}\label{21}
If $V$ is a vector space of dimension $n$, then $\dim(\P(V))=n-1$. 
If $V$ is of infinite dimension, then so is $\P(V)$.
\end{example} 
Viewing a subspace of $G$ as a projective geometry in its own right 
we may also speak of ``the dimension of a 
subspace''\index{subspace!dimension of}, which -- as to be expected 
-- is at most the dimension of $G$.
\begin{example}\label{22}
The subspaces of dimension $-1$, 0 and 1 of a projective geometry 
$G$ are respectively the empty subspace, the points and the 
projective lines. A projective geometry (or a subspace) of dimension
2 is called a {\bf projective plane}. With this terminology we may 
say that the geometry in figure \ref{333} (cf.\ \ref{19.2}) is the 
smallest irreducible projective plane: it is the so-called Fano 
plane\index{Fano plane}.
\end{example}
\par
Next we introduce some standard terminology for the projective 
geometries which are in the image of the functor 
$\P\:\Vec\to\ProjGeom$. 
\begin{definition}\label{23}
A projective geometry $G$ {\bf admits homogeneous 
coordinates}\index{homogeneous coordinates} if there exists a vector 
space $V$ such that $G\cong\P(V)$ in $\ProjGeom$.
\end{definition}
{\em For the rest of this section, we will assume all projective 
geometries to be {\bf irreducible} and to have {\bf dimension at 
least 2}}. The first condition is obviously necessary if we are to construct
homogeneous coordinates for a given projective 
geometry, cf.\ \ref{19.2}; and the latter excludes the trivial empty 
geometry, singletons and projective lines (``freak cases'', as 
E.~Artin [1957] calls them).
\par
\begin{figure}
\begin{center}
\begin{picture}(60,60)
\dashline{1}(0,30)(36,48) \dashline{1}(0,30)(60,30) 
\dashline{1}(0,30)(60,0) \dashline{1}(30,60)(30,26.25) 
\drawline(20,40)(12,24) \dashline{1}(30,60)(20,40) 
\dashline{1}(30,60)(36,48) \drawline(36,48)(60,0) 
\drawline(60,30)(20,40) \drawline(36,48)(25.7142858,30) 
\drawline(25.7142858,30)(60,0) \drawline(12,24)(60,30)
\put(25.7142858,30){\dot} \put(60,30){\dot} 
\put(20,40){\dot} \put(36,48){\dot} \put(60,0){\dot}
\put(12,24){\dot} \put(0,30){\dot}
\put(30,60){\dot}
\put(30,37.45){\dot}
\put(30,26.25){\dot}
\put(-2,32){\lett{c}} \put(10,22){\lett{a_3}} \put(63,0){\lett{b_3}} 
\put(62,32){\lett{a_2}} \put(18,42){\lett{a_1}} 
\put(38,50){\lett{b_1}} \put(23,32){\lett{b_2}}
\end{picture}
\end{center}
\caption{Desargues' property}\label{desargues}
\end{figure}
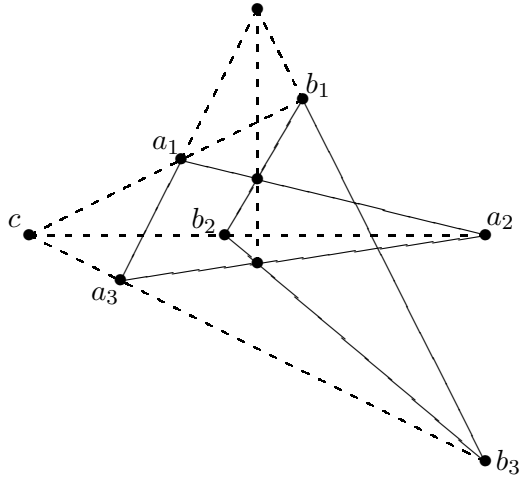

The following notion characterizes, as we will see, the 
``linearizable'' projective geometries.
\begin{definition}\label{24}
A projective geometry $G$ is {\bf arguesian}\index{projective 
geometry!arguesian}\index{arguesian!projective 
geometry|see{projective geometry}} if it is irreducible,
  has dimension at least 2, and satisfies {\bf Desargues' property}:
  for any choice of points $a_1,a_2,a_3,b_1,b_2,b_3 \in G$ for which
\begin{enumerate}
%
%
\setlength{\itemsep}{0pt} \setlength{\parskip}{0pt} 
\item there is a $c\in G$ such that $l(c,a_j,b_j)$ and $c\neq a_j \neq b_j \neq c$ hold for $j=1,2,3$,
\item no three of the points $c,a_1,a_2,a_3$ and no three of the points $c,b_1,b_2,b_3$ are collinear,
\end{enumerate}
we have that the three points $(a_i\* a_k) \cap (b_i\* 
b_k)$ with $i<k \in \{1,2,3\}$ are collinear (see figure~\ref{desargues}).
\end{definition}
As we will point out below, a theorem by R.~Baer says that 
Desargues' property guarantees the existence of certain 
automorphisms of the geometry (\ref{thmbaer}), a key result in the 
proof of the First Fundamental Theorem. It turns out that once a 
geometry is big enough, it is arguesian.
\begin{proposition}\label{25}
Every projective geometry of dimension at least $3$ (including 
$\infty$) is arguesian.
\end{proposition}
\begin{proof} See [Faure and Fr\"olicher, 2000, 8.4.6] or
  [Beutelspacher and Rosenbaum, 1998, 2.7.1].  The idea here is that
  if a projective plane is strictly embedded in another projective
  geometry, then it satisfies Desargues' property. Artin [1957, p.\ 101]
  ``suggests viewing the configuration [in figure~\ref{desargues}] as
  the projection of a three-dimensional configuration onto the plane.
  The three-dimensional configuration is easily proved [from the
  axioms].''
\end{proof}
\begin{example}
  A simple example of a projective plane that does not satisfy
  Desargues' property is the ``Moulton plane''\index{Moulton plane} (after the
  mathematician F.R.~Moulton [1902]). See e.g.\ [Beutelspacher and
  Rosenbaum, 1998, \S2.6] for a description.
\end{example}
\begin{example}
It is an exercise in linear algebra that for a vector space $V$ of 
dimension at least $3$, $\P(V)$ is arguesian [Beutelspacher and 
Rosenbaum, 1998, 2.2.1]. 
\end{example}
The content of the First Fundamental Theorem is that Desargues' 
property is also sufficient for an irreducible projective geometry 
to admit homogeneous coordinates.
\begin{theorem}[First Fundamental Theorem]\label{26}\index{projective geometry!Fundamental Theorems of}
Every arguesian projective geometry admits homogeneous coordinates.
\end{theorem}
We start the sketch of the proof by giving its general idea. We then 
proceed by a series of lemmas highlighting the key ingredients.
\begin{definition}
A {\bf hyperplane}\index{hyperplane} $H$ of a projective geometry 
$G$ is a maximal strict subspace\index{subspace!maximal strict} 
$H\subset G$: it is thus a coatom\asterix\ of $\L(G)$.
\end{definition}
The example to keep in mind for the proof of the existence of 
homogeneous coordinates is the following well-known construction of 
a projective geometry by ``adding a hyperplane at infinity'' to a 
vector space.
\begin{example}\label{23.1} [Faure and Fr\"olicher, 2000, 2.1.7]
For a vector space $V$ over $K$ we can give the disjoint union
$V\disju \P(V)$ the structure of a projective geometry using the
bijection
$$\P(V\times K) \to V \disju \P(V)\:K(x,\xi)\mapsto 
\left\{\begin{array}{l} 
\xi^{-1}x \in V\mbox{ if }\xi\neq 0 \\
Kx \in \P(V)\mbox{ if }\xi= 0
\end{array}\right. $$ 
\end{example}
Given an arguesian geometry $G$, we construct homogeneous 
coordinates $G\cong \P(V\times K) \cong V \disju\P(V)$ by 
``recovering'' the group of translations of the desired vector space 
$V$ and its group of homotheties as certain {\bf 
collineations}\index{collineation} of $G$ (i.e.\ isomorphisms from 
$G$ to itself in $\ProjGeom$) that fix a chosen hyperplane (which 
will be $\P(V)$).
\begin{definition}\label{26.1}
  A collineation $\alpha$ of $G$ is called a {\bf central
    collineation} if there is a hyperplane $H$ of $G$ (the {\bf axis}
  of $\alpha$) and a point $c \in G$ (the {\bf center} of $\alpha$)
  such that $\alpha$ fixes every point of $H$ and every line through
  $c$.
\end{definition}
Central collineations are very rigid, as the following  lemma 
shows.
\begin{lemma}\label{26.2}
  Let $\alpha$ be a central collineation of $G$ with axis $H$ and center $c$.
  Let $p$ be a point in $G\setminus(\{c\}\cup H)$. Then $\alpha$ is uniquely
  determined by $\alpha(p)$. In particular, for every $x \in G$ not on
  $H$ nor on $c \* p$, we have
  \begin{equation}\alpha(x) = (c \* x) \cap (f \*\alpha(p)) \label{coll}
  \end{equation}
   where $f\defeq(p\* x) \cap H$.
\end{lemma}
\begin{proof} See [Beutelspacher and Rosenbaum, 1998, 3.1.3] Note that
  $\alpha(x) \in c\* x$ because $\alpha$ has center $c$ and $\alpha(x)
  \in f\*\alpha(p)$ because $x\in p\* x$ and $\alpha$ has axis $H$.
  These two lines intersect in a point because they are distinct (as
  $x \not\in c\* p = c\*\alpha(p)$) and they lie in the plane spanned
  by $p,x$ and $c$.
\end{proof}  
\par
Here is the announced theorem where Desargues' property comes into 
play.
\begin{lemma}[Baer's existence theorem of central collineations]\label{thmbaer}\index{collineation!Baer's existence theorem for central}
 Let $G$ be arguesian. If $H$ is a hyperplane and $c,p,p'$ are
 distinct collinear points of $G$ with $p,p' \not\in H$, then there
 is exactly one collineation of $G$ with center $c$ and axis $H$
 mapping $p$ to $p'$.
\end{lemma}
\begin{proof} See [Beutelspacher and Rosenbaum, 1998, 3.1.8] or [Faure and 
Fr\"olicher, 2000, 8.4.11]. The basic idea is to use (\ref{coll}) to 
define the map. Desargues' property is then used to show that it is 
well-defined and a collineation, through rather lengthy geometric 
verifications.
\end{proof}
We now fix a hyperplane $H$ of $G$ and a point $o \in G\backslash 
H$. Let $T$ be the set of collineations with axis $H$ and center 
{\it on} $H$. We call an element of $T$ a {\bf 
translation}\index{collineation!translation}.
\begin{lemma}
$T$ is an abelian group (under composition) which acts simply 
transitively on $G\setminus H$. Translating this action of $T$ into 
an addition on $V\defeq G\setminus H$, the latter also becomes an 
abelian group.
\end{lemma}
\begin{proof} [Beutelspacher and Rosenbaum, 1998, 3.2.2] Simple 
transitivity of the action means that if $p\neq q$ in $G\setminus H$ 
then there is a unique collineation of $G$ with axis $H$ and center 
$(p\* q)\cap H$ sending $p$ to $q$; it is a consequence 
of~\ref{thmbaer}. The fact that $T$ is a subgroup of the group of 
collineations uses the fact that if a collineation has an axis, it 
has a center, and the rigidity of lemma~\ref{26.2}. The 
commutativity of the group requires a little more work. 
\par
The simple transitivity allows us to transport the group structure 
from $T$ to $V$. Indeed, for $p \in V$, denote $\tau_p$ the unique 
element in $T$ such that $\tau_p(o)=p$. For $p,q \in V$, we then put
$$p+q \defeq \tau_p(q) = \tau_p(\tau_q(o)).$$
\end{proof}
Next, let $K^{\times}$ be the group (under composition) of all 
collineations of $G$ with center $o$ and axis $H$. We call an 
element of $K^{\times}$ a {\bf 
homothety}\index{collineation!homothety}. It is an immediate 
consequence of~\ref{thmbaer} that $K^{\times}$ acts simply 
transitively on $L\setminus\{o\}$ for every line $L$ through $o$. Let 
$\alpha_o$ be the constant morphism $G\to G\: p \mapsto o$. 
\begin{lemma}
If on the set $K\defeq K^{\times} \cup \{\alpha_o\}$, we define 
addition by
\[(\sigma_1 +\sigma_2) (x) \defeq \sigma_1(x) + \sigma_2(x) \textrm{ for every }x\in V\]
and multiplication by composition, then $K$ becomes a field.
\end{lemma}
\begin{proof} [Beutelspacher and Rosenbaum, 1998, 3.3.4] The main 
difficulty here is showing that $K$ is closed under this addition. 
\end{proof}
This multiplication is not commutative in general. {\it 
Pappus' Theorem}\index{Pappus' Theorem} geometrically characterizes 
those arguesian geometries for which it is (see [Artin, 1957, 
chapter II, \S7] for example). 
\begin{lemma}
The action of $K$ on $V$ by 
$$K\times V\to V\:(\sigma,x)\mapsto\sigma\cdot x\defeq\sigma(x)$$
is a ``scalar multiplication'' making $V$ a (left) vector space over $K$.
\end{lemma}
\begin{proof} 
[Beutelspacher and Rosenbaum, 1998, 3.3.5] Showing that 
$\sigma(-x)=-\sigma(x)$ is what requires the most (but not that 
much) work. 
\end{proof}
The next lemma then finishes the proof of~\ref{26}.
\begin{lemma}
$G$ is isomorphic as a projective geometry to $\P(V \times K)$. 
\end{lemma}
\begin{proof} [Beutelspacher and Rosenbaum, 1998, 3.4.2] Remember that 
$V=G\backslash H$. The isomorphism $\phi\: G \to \P(V\times K)$ is 
defined by
\[\phi(x) \defeq 
\left\{ \begin{array}{l} 
K(x,1)\mbox{ if }x\in G\backslash H \\ %
K(y,0)\mbox{ if }x\in H \textrm{, where }y\neq x \textrm{ is any 
point of } o\* x
\end{array}\right. \] 
which identifies $H$ with $\P(V)$ as expected.
\end{proof}
\par
Having dealt with objects, we move on to the linear representation 
of (some of) the morphisms of $\ProjGeom$. Cl.-A.~Faure [2002] has 
provided a short and elementary proof of the next theorem, which 
originally appeared in [Faure and Fr\"olicher, 1994]. 
\begin{theorem}[Second Fundamental Theorem]\label{27}\index{projective geometry!Fundamental Theorems of} Let $V_1$ and $V_2$ be vector spaces.
  Every {\bf non-degenerate morphism}\index{morphism!of projective
    geometries}\index{non-degenerate morphism|see{morphism of
      projective geometries}} $g\:\P(V_1)\parto\P(V_2)$, meaning that
  its image contains three non-collinear points, is of the form
  $\P(f)$ for some semilinear map $f\:V_1\shorterto V_2$. Moreover $f$ is
  unique up to scalar multiplication.
\end{theorem}
The uniqueness is an immediate consequence of the following fact (see
e.g.\ [Faure, 2002, 2.4]), which we record here for future reference.
Its proof is an exercise in linear algebra.
\begin{proposition}\label{27.1}
Let $f_1, f_2\: V_1 \to V_2$ be two additive maps between vector 
spaces $(K_1,V_1)$ and $(K_2,V_2)$. Assume that $f_2(x)\in K_2 
f_1(x)$ for every $x \in V_1$ and that $f_1(V_1)$ contains two 
linearly independent vectors. Then there exists a $\mu \in K_2$ 
such that $f_2 = \mu f_1$.
\end{proposition}
The Second Fundamental Theorem implies that the vector space whose 
existence was guaranteed by the First Fundamental Theorem is 
essentially unique.
\begin{corollary}\label{abc}
  If $\phi\:\P(V_1)\to\P(V_2)$ is an isomorphism in $\ProjGeom$ where
  $V_1$ is a $K_1$-vector space of dimension at least $3$ and $V_2$ is a
  $K_2$-vector space (with $K_1,K_2$ fields), then there exists a field
  isomorphism $\sigma\: K_1 \to K_2$ and a bijective $\sigma$-linear map
  $f\: V_1 \to V_2$ such that $\phi = \P(f)$.
\end{corollary}
Remark that the uniqueness of homogeneous coordinates holds for 
projective dimension at least $2$ while existence needs dimension at 
least $3$ (or Desargues' property).
\par
A composition of two non-degenerate morphisms need not be 
non-degenerate (think of the composition $f \circ g$ of two linear 
maps with $\im g \subseteq \ker f$). A morphism of projective 
geometries is called {\bf arguesian}\index{morphism!of projective 
geometries}\index{arguesian!morphism|see{morphism of projective 
geometries}} when it is the composite of finitely many 
non-degenerate morphisms. The following proposition [Faure and 
Fr\"o\-lich\-er, 2000, 10.3.1] says that these are exactly the morphisms 
induced by semilinear maps.
\begin{proposition}
For a partial map $g: \P(V_1) \parto \P(V_2)$ between arguesian
geometries, the following conditions are equivalent:
\begin{enumerate}
%
%
\setlength{\itemsep}{0pt} \setlength{\parskip}{0pt} 
\item\label{arg1} $g$ is induced by a semilinear map $f\: V_1 \to V_2$, 
\item\label{arg2} $g$ is the composite of two non-degenerate morphisms between arguesian geometries, 
\item\label{arg3} $g$ is the composite of finitely many non-degenerate morphisms between arguesian geometries.
\end{enumerate}
\end{proposition}
\begin{proof}
The only nontrivial implication is (\ref{arg1} $\Rightarrow$ \ref{arg2}). Let $\sigma$ be the field homomorphism associated to $f$. By hypothesis $\dim(V_2)\geq 3$  and we can pick three linearly independent vectors $y_1, y_2, y_3 \in V_2$. Put $W\:= V_1 \times K_1^3$. Now define the maps 
\begin{eqnarray*}
& &f_1\: V_1 \to W\: x\mapsto (x,0,0,0) \\
& &f_2\:W \to V_2:(x,k_1,k_2,k_3)\mapsto f(x) +\sigma(k_1)y_1+\sigma(k_2)y_2+\sigma(k_3)y_3
\end{eqnarray*}
Then $f=f_2\circ f_1$ and thus $g=\P(f)=\P(f_2) \circ\P(f_1)$, where $\P(f_1)$ and $\P(f_2)$ are clearly non-degenerate.
\end{proof}
We define the category\index{projective geometry!category of 
arguesian -ies}\index{projective geometry!arguesian} $\sf Arg$ of 
arguesian projective geometries and arguesian morphisms. The 
Fundamental Theorems may then be summarized in the following 
statement, which is as powerful as one could hope.
\begin{theorem}\label{28}
The functor $\P\:\SemiVec_{\dim\geq 3}\to\mathsf{Arg}$ is 
essentially surjective and essentially injective on objects, full, 
and only identifies semilinear maps when they are a nonzero scalar 
multiple of each other.\index{projective geometry!Fundamental 
Theorems of}
\end{theorem}

\section{Hilbert geometries, Hilbert lattices, propositional systems}\label{E}
\setcounter{theorem}{0}
A (real, complex, quaternionic or generalized) Hilbert space 
$H$\index{generalized Hilbert space} is in particular a vector space, so by 
\ref{8} its one-dimensional linear subspaces form a projective 
geometry $\P(H)$. But the orthogonality relation on the elements of 
$H$, defined as $x\perp y$ if and only if the inner product of $x$ 
and $y$ is zero, obviously induces an {\it orthogonality relation} 
on $\P(H)$: $A\perp B$ in $\P(H)$ when $a\perp b$ for some $a\in 
A\setminus\{0\}$ and $b\in B\setminus\{0\}$. We make an abstraction 
of this.
\begin{definition}\label{100}
Given a binary relation $\perp\ \subseteq G\times G$ on a projective 
geometry $G$ and a subset $A\subseteq G$, we put 
$A\ortho\defeq\{b\in G\mid \forall a\in A:b\perp a\}$. A {\bf 
Hilbert geometry}\index{Hilbert geometry} $G$ is a projective 
geometry together with an {\bf orthogonality 
relation}\index{orthogonality relation} $\perp\ \subseteq G\times G$ 
such that, for all $a,b,c,p\in G$,
\begin{enumerate}
%
%
\setlength{\itemsep}{0pt} \setlength{\parskip}{0pt} 
\item[(O1)] if $a\perp b$ then $a\neq b$,
\item[(O2)] if $a\perp b$ then $b\perp a$,
\item[(O3)] if $a\neq b$, $a\perp p$, $b\perp p$ and $l(a,b,c)$ then $c\perp p$,
\item[(O4)] if $a\neq b$ then there is a $q\in G$ such that $l(q,a,b)$ and $q\perp a$,
\item[(O5)] if $S\subseteq G$ is a subspace such that $S\dblortho=S$, then $S\vee S\ortho=G$.
\end{enumerate}
\end{definition}
Very often we shall simply speak of a ``Hilbert geometry $G$'', 
leaving both the collinearity $l$ and the orthogonality $\perp$ 
understood. A subspace $S\subseteq G$ is said to be {\bf 
(biorthogonally) closed}\index{subspace!closed} if $S\dblortho=S$. 
\par
Axioms (O1--4) in the above definition say in particular that a 
Hilbert geometry is a `state space'\index{state space} in the sense 
of [Moore, 1995] as we explain in \ref{82}. The fifth axiom could 
have been written as: $S=S\dblortho$ if and only if $S\vee 
S\ortho=G$, because (as we shall show in \ref{104} (\ref{g}) in a 
more abstract setting) the necessity is always true. We make some 
more comments on these axioms in section \ref{H}.
\par
The term `Hilbert geometry' is well-chosen, as C. Piron's now famous 
example [1964, 1976] shows.
\begin{definition}\label{ghs}
A {\bf generalized Hilbert space}\index{generalized Hilbert space} 
(also called {\bf orthomodular space}\index{orthomodular 
space|see{generalized Hilbert space}}) $(H,K,{\ }^*,\inprod{\ ,\ })$ 
is a vector space $H$ over a field $K$ together with an involutive 
anti-automorphism $K\to K\:\alpha\mapsto\alpha^*$ and an {\bf 
orthomodular Hermitian form}\index{Hermitian form} $H\times 
H\to K\:(x,y)\mapsto\inprod{x,y}$, that is, a form satisfying
\begin{enumerate}
%
%
\setlength{\itemsep}{0pt} \setlength{\parskip}{0pt} 
\item[(S1)] $\inprod{\lambda x_1 +x_2, y} = \lambda\inprod{x_1,y}+\inprod{x_2,y}$ for all $x_1,x_2,y\in H, \lambda \in K$,
\item[(S2)] $\inprod{y,x} = \inprod{x,y}^*$ for all $x,y \in H$,
\end{enumerate}
 and such that, when putting 
$S\ortho\defeq\{x\in H\mid\forall y\in S:\inprod{x,y}=0\}$ for a 
linear subspace $S\subseteq H$, 
\begin{enumerate} 
\item[(S3)] $S=S\dblortho$ implies $S\oplus S\ortho=H$.
\end{enumerate}
\end{definition}
Note that an orthomodular Hermitian form is automatically {\bf 
anisotropic},
\begin{enumerate}
\item[(S4)] $\inprod{x,x} \neq 0$ for all $x \in H\setminus\{0\}$,
\end{enumerate} and that in the finite dimensional case the converse
is true too.  I.~Amemiya and H.~Araki [1966] proved that when $K$ is
one of the ``classical'' fields equipped with its ``classical''
involution ($\RR$ with identity, $\CC$ and $\HH$ with their respective
conjugations), the definition of `generalized Hilbert space' is
equivalent to the ``classical'' definition of a Hilbert space as
inner-product space which is complete for the metric induced by the
norm. While H.~Keller [1980] was the first to construct a
``nonclassical'' generalized Hilbert space, M.~Sol\`er [1995] proved
that an infinite dimensional generalized Hilbert space $H$ is
``classical'' precisely when $H$ contains an orthonormal
sequence.\index{generalized Hilbert space!Sol\`er's Theorem} We refer
to [Holland, 1995] for a nice survey, and to A. Prestel's [2006]
contribution to this handbook for a complete and historically
annotated proof of Sol\`er's theorem. For a comment on the
lattice-theoretic meaning of Sol\`er's theorem, see \ref{86}.
\begin{example}\label{101}
For a generalized Hilbert space $H$, the projective geometry $\P(H)$ 
together with the obvious orthogonality relation forms a Hilbert 
geometry\index{Hilbert geometry!induced by a generalized Hilbert 
space}: axioms (O1--3) are immediate, (O4) follows from a 
standard Gram-Schmidt trick and (O5) is also immediate since $S\vee 
S^{\perp}=S\oplus S^{\perp}$ for any linear subspace $S$ of $H$.
\end{example}
\par
From our work in section \ref{B} we know that, since a Hilbert 
geometry is in particular a projective geometry, the lattice of 
subspaces $\L(H)$ is a projective lattice. Because of the 
orthogonality relation on $G$, there is some extra structure on 
$\L(G)$; the following proposition identifies it.
\begin{proposition}\label{102}
If $G$ is a Hilbert geometry with orthogonality relation $\perp$, 
then the operator $\mbox{ }\ortho\:\L(G)\to\L(G)\:S\mapsto S\ortho$ 
satisfies, for all $S,T\in\L(G)$,
\begin{enumerate}
%
%
\setlength{\itemsep}{0pt} \setlength{\parskip}{0pt} 
\item $S\subseteq S\dblortho$,
\item if $S\subseteq T$ then $T\ortho\subseteq S\ortho$,
\item $S\cap S\ortho=\emptyset$,
\item if $S=S\dblortho$ and $a\in G$ then $\{a\}\vee S=(\{a\}\vee S)\dblortho$.
\item if $S=S\dblortho$ then $S\vee S\ortho=G$.
\end{enumerate}
\end{proposition}
\begin{proof} All is straightforward, except for (iv). We need to prove 
that $(\{a\}\vee S)\dblortho\subseteq\{a\}\vee S$ for 
$S=S\dblortho$. If $a\in S$ then this is trivial so we suppose from 
now on that $a\not\in S=S\dblortho$, i.e.\ there exists $p\in 
S\ortho$ such that $a\not\perp p$. Let $b\in(\{a\}\vee S)\dblortho$; 
if $b=a$ or $b\in S$ then obviously $b\in\{a\}\vee S$. If $b\neq a$ 
and $b\not\in S$ then we claim that $(a\* b)\cap\{p\}\ortho$ is a 
singleton and moreover that its single element, call it $q$, belongs 
to $S$. This then proves the assertion, for $q\perp p$ implies 
$q\neq a$, which makes $q\in a\*b$ imply that $b\in a\*q\subseteq 
\{a\}\vee S$.
\par
Now $(a\* b)\cap\{p\}\ortho$ is non-empty\footnote{What we {\it 
really} prove here is that for $a,b,p$ with $a\neq b$ in a Hilbert 
geometry $G$ there always exists some $q\in a\* b$ such that $q\perp 
p$. This statement, which is obviously stronger than (O4), is often 
used instead of (O4). See \ref{82.0} for a relevant comment.}, 
because in case that $a\not\perp p\not\perp b$ we can always pick 
$x\in p\*a$ and $y\in p\*b$ such that $x, y\in\{p\}\ortho$ by (O4); 
then $x\neq a$ and $y\neq b$ so $p\in(a\*x)\cap(b\*y)$ and (G3) thus 
gives a $q\in (a\*b)\cap(x\*y)\subseteq(a\*b)\cap\{p\}\ortho$ (for 
$\{p\}\ortho$ is a subspace by (O3)). Would $q_1\neq q_2\in(a\* 
b)\cap\{p\}\ortho$, then $l(q_1,q_2,a)$ by (G2) hence 
$a\in\{p\}\ortho$, a contradiction. So we conclude that $(a\* 
b)\cap\{p\}\ortho=\{q\}$.
\par
We shall show that $q\in S=S\dblortho$, i.e.\ for any $r\in S\ortho$ 
we have $q\perp r$. For $r=p$ this is true by construction; for 
$r\neq p$ we may determine, by the ``same'' argument as above, a 
(unique) point $s\in\{a\}\ortho\cap(p\* r)\subseteq(\{a\}\ortho\cap 
S\ortho)=(\{a\}\vee S)\ortho$. The latter equality can be shown with 
a simple calculation, but we also give a more abstract proof in 
\ref{104} (\ref{f}). Because $a,b\in(\{a\}\vee S)\dblortho$ it 
follows that $a\perp s$ and $b\perp s$; hence we get $q\perp s$ from 
$q\in a\*b$ and (O3). But also $s\neq p$ follows, thus $s\in p\* r$ 
implies $r\in p\*s$, and because we know that $q\perp p$ too, we 
finally obtain $q\perp r$, again from (O3).
\end{proof}
This proposition calls for a new definition.
\begin{definition}\label{103}
A projective lattice $L$ is a {\bf Hilbert lattice}\index{Hilbert 
lattice} if it comes with an {\bf orthogonality operator} $\mbox{ 
}\ortho\:L\to L\:x\mapsto x\ortho$ satisfying, for all $x,y\in L$,
\begin{enumerate}
%
%
\setlength{\itemsep}{0pt} \setlength{\parskip}{0pt} 
\item[(H1)] $x\leq x\dblortho$,
\item[(H2)] if $x\leq y$ then $y\ortho\leq x\ortho$,
\item[(H3)] $x\wedge x\ortho=0$,
\item[(H4)] if $x=x\dblortho$ and $a$ is an atom of $L$ then $a\vee x=(a\vee x)\dblortho$,
\item[(H5)] if $x=x\dblortho$ then $x\vee x\ortho=1$.
\end{enumerate}
\end{definition}
Usually we shall simply speak of ``a Hilbert lattice $L$'', and 
leave the orthogonality operator understood.
\par
The crux of \ref{102} is thus that the projective lattice of 
subspaces of a Hilbert geometry is a Hilbert lattice. Having \ref{9} 
in mind, it should not come as a surprise that there is a converse 
to this. However, we shall not give a direct proof of such a 
statement, for we wish to involve yet another mathematical 
structure. Again the source of inspiration is the concrete example 
of Hilbert spaces: the subspaces $S\subseteq H$ for which 
$S=S\dblortho$ are particularly important, for they are precisely 
the subspaces which are closed for the norm topology on $H$ 
(see [Schwartz, 1970, p.~392] for example). Also in the abstract case they are worth 
a closer look.
\par
By a {\bf (biorthogonally) closed element} of a Hilbert lattice 
$(L,\mbox{ }\ortho)$ we shall of course mean an $x\in L$ for which 
$x=x\dblortho$. We write $\C(L)\subseteq L$ for the ordered set of 
closed elements, with order inherited from $L$. We shall now discuss 
some features of this ordered set that -- as it will turn out -- 
describe it completely. First we prove a technical lemma.
\begin{lemma}\label{104}
For a Hilbert lattice $L$,
\begin{enumerate}
%
%
\setlength{\itemsep}{0pt} \setlength{\parskip}{0pt} 
\item\label{c} if $x\in L$ then $x\ortho\in\C(L)$,
\item\label{a} $0\dblortho=0$, $0\ortho=1$ and $1\ortho=0$,
\item\label{f} $(\bigvee_i x_i)\ortho=\bigwedge_ix_i\ortho$ for $(x_i)_{i\in I}\in L$, 
\item\label{g} if $x\vee x\ortho=1$ then $x=x\dblortho$,
\item\label{d} the map $\mbox{ }\dblortho\:L\to L\:x\mapsto x\dblortho$ is a closure operator with fixpoints $\C(L)$.
\end{enumerate}
\end{lemma}
\begin{proof} Statements (\ref{c}) and (\ref{a}) are almost trivial. For 
(\ref{f}) one uses (H1--2) over and again to verify $\geq$ and 
$\leq$ as follows:
\begin{center}
\begin{tabular}{lc|cl}
$\forall k\in I: \bigwedge_ix_i\ortho\leq x_k\ortho$ & & & $\forall k\in I: x_k \leq \bigvee_ix_i$ \\
$\Rightarrow \forall k\in I: x_k\leq x_k\dblortho\leq(\bigwedge_ix_i\ortho)\ortho$ & & & $\Rightarrow \forall k\in I: (\bigvee_ix_i)\ortho\leq x_k\ortho$ \\
$\Rightarrow \bigvee_ix_i\leq (\bigwedge_ix_i\ortho)\ortho$ & & & $\Rightarrow (\bigvee_ix_i)\ortho\leq \bigwedge_ix_i\ortho$ \\
$\Rightarrow \bigwedge_ix_i\ortho\leq(\bigwedge_ix_i\ortho)\dblortho\leq(\bigvee_ix_i)\ortho$ & & &  \\
\end{tabular}
\end{center}
As for (\ref{g}), the assumption together with (H3) and modularity 
in $L$ (for $x\leq x\dblortho$) give $x=x\vee 0=x\vee(x\ortho\wedge 
x\dblortho)=(x\vee x\ortho)\wedge x\dblortho=1\wedge 
x\dblortho=x\dblortho$. Finally, it straightforwardly follows from 
(H2) that 
$$\phi\:L\to\C(L)\:x\mapsto x\dblortho\mbox{ \ and \ }\psi\:\C(L)\to L\:y\to y$$ 
are maps that preserve order, and they satisfy $\phi(x)\leq y\iff x\leq 
\psi(y)$ for any $x\in L$ and $y\in\C(L)$. So these maps are 
adjoint, $\phi\dashv\psi$, and since moreover $\phi$ is surjective and $\psi$ injective, the 
composition $\psi\circ\phi\:L\to L\:x\mapsto x\dblortho$ is a 
closure operator with fixpoints $\C(L)$, as claimed in (\ref{d}).
\end{proof}
For closed elements $(x_i)_{i\in I}\in L$ we shall write 
$\clbigvee_i x_i$ for $(\bigvee_i x_i)\dblortho$, and in particular 
$x\clvee y$ for $(x\vee y)\dblortho$.
\begin{proposition}\label{105}
For any Hilbert lattice $L$, the ordered set $(\C(L),\leq)$ together 
with the restricted operator $\mbox{ 
}\ortho\:\C(L)\to\C(L)\:x\mapsto x\ortho$ is a complete, atomistic, 
ortho\-mod\-ular\asterix\ \index{modular 
lattice!ortho-}\index{orthomodular lattice|see{modular lattice}} 
lattice satisfying the covering law.
\end{proposition}
\begin{proof} By (\ref{d}) of \ref{104}, $\C(L)$ is a complete lattice 
inheriting infima from $L$, and with suprema given by $\clbigvee$. 
Moreover, (H2--3) assert that $x\mapsto x\ortho$ is an 
ortho\-complem\-ent\-ation\asterix\ on $\C(L)$. It is 
straightforward from (H4) and \ref{104} (\ref{a}) that the atoms of 
$L$ are closed; and conversely is it clear that the atoms of $\C(L)$ 
are atoms of $L$ too. So $\C(L)$ is atomistic, because $L$ is. In 
the same way, since $L$ has the covering law (cf.\ \ref{9.1}) and 
the atoms of $L$ are precisely those of $\C(L)$, again (H4) assures 
that $\C(L)$ has the covering law too. Finally, if $x\leq y$ in 
$\C(L)$ then by the modular law in $L$ and (H5)
$$x\clvee(x\ortho\wedge y)=(x\vee(x\ortho\wedge y))\dblortho=((x\vee x\ortho)\wedge y)\dblortho=(1\wedge y)\dblortho=y\dblortho=y;$$ i.e.\ the orthomodular law holds in $\C(L)$.
\end{proof}
\begin{example}\label{105.1}
By (H4) it follows that, if $a_1,...,a_n$ are atoms of a Hilbert 
lattice $L$, then (each one of them is closed and) 
$a_1\clvee...\clvee a_n=a_1\vee...\vee a_n$. If $L$ is a Hilbert 
lattice\index{Hilbert lattice!of finite rank} of finite rank, then 
every $x\in L$ can be written as a {\it finite} supremum of atoms 
(this is true for any atomistic lattice satisfying the covering law 
of finite rank, see e.g.\ [Maeda and Maeda, 1970, section 8]), hence 
$x=x\dblortho$; i.e.\ $L\cong\C(L)$. So if $G$ is a Hilbert 
geometry\index{Hilbert geometry!of finite dimension} of finite 
dimension, then every subspace of $G$ is biorthogonally closed; in 
particular is this true for $\P(H)$ when $H$ is a (generalized) 
Hilbert space of finite dimension.
\end{example}
\par
Inspired by the result in \ref{105} we now give another definition 
due to C.\ Piron [1964, 1976]. 
\begin{definition}\label{106}
An ordered set $(C,\leq)$ with an operator $\mbox{ }\ortho\:C\to 
C\:x\mapsto x\ortho$ is a {\bf propositional 
system}\index{propositional system} if it is a complete, atomistic, 
orthomodular lattice that satisfies the covering law\index{covering 
law} (with $x\mapsto x\ortho$ as orthocomplementation).
\end{definition}
We shall speak of ``a propositional system $C$'', always using 
$x\ortho$ as notation for the orthocomplement of $x\in C$. And we 
shall continue to write $\clbigvee_ix_i$ for the supremum in $C$, 
and $\bigwedge_ix_i$ for the infimum.
\begin{example}\label{107}
The closed subspaces of a generalized Hilbert space $H$ form a 
propositional system\index{propositional system!induced by a 
generalized Hilbert space}, that we shall write as $\C(H)$ instead 
of $\C(\L(H))$.
\end{example}
\par
According to \ref{105} and \ref{106}, the closed elements of a 
Hilbert lattice form a propositional system. Earlier we proved (cf.\ 
\ref{102} and \ref{103}) that the subspaces of a Hilbert geometry 
form a Hilbert lattice. It is now time to come full circle: we want 
to associate a Hilbert geometry to a given propositional system. The 
lattice-theoretical results in \ref{9.1} will be useful here too.
\begin{proposition}\label{108}
The set $\G(C)$ of atoms of a propositional system $C$ form a 
Hilbert geometry\index{Hilbert geometry!induced by a propositional 
system} for collinearity and orthogonality given by
\begin{equation}\label{109}
l(a,b,c)\mbox{ if }a\leq b\clvee c\mbox{ or }b=c,\hspace{3ex}a\perp 
b\mbox{ if }a\leq b\ortho.
\end{equation}
\end{proposition}
\begin{proof} By definition $C$ is complete, atomistic and satisfies the 
covering law; therefore it is upper semimodular by (\ref{ii}) of 
\ref{9.1}. But $C$ is also orthocomplemented\asterix, so it is 
isomorphic to its opposite\asterix\ (by $C\to C\op\:x\mapsto 
x\ortho$): upper semimodularity thus implies lower semimodularity. 
Then $C$ must have the intersection property by (\ref{iii}) of 
\ref{9.1}, and so its atoms form a projective geometry for the 
indicated collinearity.
\par
Now we must check the axioms for the orthogonality relation; the 
first three are (almost) trivial. For (O4), if $a\neq b$ in $\G(C)$ 
then $b\leq a\clvee a\ortho=1$ hence, by the intersection property, 
$(a\clvee b)\wedge a\ortho\neq 0$; the atomisticity of $C$ gives us 
a $q\in\G(C)$ such that $q\leq a\clvee b$ and $q\leq a\ortho$, as 
wanted. Finally, (O5) requires some more sophisticated calculations. 
First note that, for any subspace $S\subseteq\G(C)$ and element 
$a\in\G(C)$,
$$a\in S\ortho\iff\forall b\in S: b\leq a\ortho\iff\clbigvee S\leq a\ortho\iff a\leq(\clbigvee S)\ortho.$$
Thus we always have that $S\ortho=\{a\in\G(C)\mid a\leq(\clbigvee 
S)\ortho\}$, which by atomisticity of $C$ means that 
$\clbigvee(S\ortho)=(\clbigvee S)\ortho$; in particular is $S$ 
closed, $S=S\dblortho$, if and only if $S=\{a\in\G(C)\mid 
a\leq\clbigvee S\}$. If $S$ is a trivial subspace, then it is clear 
that $S\vee S\ortho=\G(C)$; so from now on, let $S=S\dblortho$ be 
non-trivial. By the projective law, valid in $\L(\G(C))$ as in any 
other projective lattice, $S\vee S\ortho=\G(C)$ just means that for 
any $p\in\G(C)$ there exist $a,b\in \G(C)$ such that $a\leq\clbigvee 
S$, $b\leq(\clbigvee S)\ortho$ and $p\leq a\clvee b$. And this is 
indeed true in the propositional system $C$; to simplify notations 
we shall write $x\defeq\clbigvee S$ in the argument that 
follows\footnote{This argument {\it actually} shows that for any 
$p\in\G(C)$ and any $x\in C$ which is not $0$ nor $1$, there exist 
$a,b\in\G(C)$ such that $a\leq x$, $b\leq x\ortho$ and $p\leq 
a\clvee b$; see also [Maeda and Maeda, 1970, 30.7].}. Suppose first 
that $x\wedge(x\ortho\clvee p)\not\leq x\ortho$, then (by $C$'s 
atomisticity) there must exist an $a\in\G(C)$ such that $a\leq 
x\wedge(x\ortho\clvee p)$ and $a\not\leq x\ortho$. If $a=p$ then 
$p\leq x$ and we can pick any atom $b\leq x\ortho$ to show that 
$p\leq a\clvee b$ as wanted. If $a\neq p$ then from $a\leq 
x\ortho\clvee p$ and the intersection property we get an atom $b\leq 
x\ortho\wedge(a\clvee p)$; but certainly is $a\neq b$ (because 
$a\leq x$ and $b\leq x\ortho$) so $b\leq a\clvee p$ is equivalent to 
$p\leq a\clvee b$ by \ref{9.1} (\ref{iv}), as wanted. Next suppose 
that $x\wedge(x\ortho\clvee p)\leq x\ortho$; this means that 
$x\ortho=x\ortho\clvee(x\wedge(x\ortho\clvee p))=x\ortho\clvee p$ by 
orthomodularity in the second equality, so $p\leq x\ortho$. Picking 
any atom $a\leq x$ and putting $b\defeq p$ we have $p\leq a\clvee b$ 
as wanted.
\end{proof}
Note that, for a given Hilbert lattice $L$, the atoms of $\C(L)$ are 
exactly those of $L$, and the supremum of two atoms in $\C(L)$ is 
equal to their supremum in $L$. Thus it follows that $\G(\C(L))$ (as 
in \ref{108}) is the \textit{same} projective geometry as $\G(L)$ 
(as in \ref{9}).
\par
Our aim is to build a \textit{triple} categorical equivalence 
between Hilbert geometries, Hilbert lattices and propositional 
systems. We must therefore define an appropriate notion of `morphism 
between propositional systems'. And then it turns out that we must 
restrict the morphisms between Hilbert geometries, resp.\ Hilbert 
lattices, if we want to establish such a triple equivalence.
\begin{definition}\label{34}
Let $C_1$ and $C_2$ be propositional systems. A map $h\:C_1\to C_2$ 
is a {\bf morphism of propositional systems}\index{morphism!of 
propositional systems} if it preserves arbitrary suprema and maps 
atoms of $C_1$ to atoms or the bottom element of $C_2$.
\end{definition}
It is a simple observation that propositional systems and their 
morphisms form a category $\PropSys$.
\par
We shall now adapt the definition of `morphism' between Hilbert 
geometries, resp.\ Hilbert lattices: since these structures come 
with their respective closure operators, it is natural to consider 
`continuous morphisms'.
\begin{definition}\label{31}
Let $G_1$ and $G_2$ be Hilbert geometries. A morphism of projective 
geometries $g\:G_1\parto G_2$ (as in \ref{11}) is {\bf continuous} 
when, for every closed subspace $F$ of $G_2$, $g^*(F)$ is a closed 
subspace of $G_1$.
\par
Let $L_1$ and $L_2$ be Hilbert lattices. A morphism of projective 
lattices $f\:L_1\to L_2$ (as in \ref{13}) is {\bf 
continuous}\index{morphism!of projective lattices} when 
$f(x\dblortho)\leq f(x)\dblortho$ for every $x\in L_1$.
\end{definition}
Hilbert geometries and continuous morphisms form a
category\index{Hilbert geometry!category of -ies} $\HilbGeom$, and
there is a faithful functor $\HilbGeom\to\ProjGeom$ that ``forgets'' 
the orthogonality relation on a Hilbert geometry. Similarly Hilbert 
lattices and continuous morphisms form a category $\HilbLat$ with a 
forgetful functor to $\ProjLat$.
\begin{example}\label{31.1}
We say that a semilinear map $f\:H_1\to H_2$ between generalized 
Hilbert spaces is {\bf continuous}\index{semilinear map} when it is 
so in the usual sense of the word with respect to biorthogonal 
closure. This is precisely saying that the induced morphism 
$\P(f)\:\P(H_1)\parto\P(H_2)$ is continuous in the sense of 
\ref{31}. There is then a category\index{generalized Hilbert 
space!category of -s} $\GenHilb$ of generalized Hilbert spaces and 
continuous semilinear maps, and also a functor
$$\P\:\GenHilb\to\HilbGeom\:\Big(f\:H_1\to H_2\Big)\mapsto\Big(\P(f)\:\P(H_1)\parto\P(H_2)\Big).$$
The Second Fundamental Theorem \ref{27} implies that every 
non-degenerate continuous morphism between Hilbert geometries 
$\P(H_1)$ and $\P(H_2)$ is induced by a continuous semilinear map.
\end{example}
In passing we note that the forgetful functor 
$\HilbGeom\to\ProjGeom$ is not full: there exist non-continuous 
linear maps between Hilbert spaces, and these induce non-continuous 
morphisms between (the underlying projective geometries of) Hilbert 
geometries.
\par
The following is then the expected amendment of \ref{14}.
\begin{proposition}\label{35}
If $g\:G_1\parto G_2$ is a continuous morphism between Hilbert 
geometries then $$\L(g)\:\L(G_1)\to\L(G_2)\:S\mapsto 
\bigcap\{T\in\L(G_2)\mid S\subseteq g^*(T)\}$$ is a continuous 
morphism between Hilbert lattices. If $f\:L_1\to L_2$ is a 
continuous morphism between Hilbert lattices then its restriction to 
closed elements
$$\C(f)\:\C(L_1)\to\C(L_2)\:x\mapsto f(x)\dblortho$$ 
is a morphism of propositional systems. And if $h\:C_1\to C_2$ is a 
morphism between propositional systems then 
$$\G(h)\:\G(C_1)\parto\G(C_2)\:a\mapsto h(a)\mbox{ with kernel }\{a\in\G(C_1)\mid h(a)=0\}$$
is a continuous morphism between Hilbert geometries.
\end{proposition}
\begin{proof} For a continuous morphism $g\:G_1\parto G_2$ between Hilbert 
geometries and $S\in\L(G_1)$ we can compute, with notations as in 
\ref{14}, that
\begin{eqnarray*}
 & & S\subseteq g^*\Big(\L(g)(S)\Big)\subseteq g^*\Big(\Big(\L(g)(S)\Big)\dblortho\Big) \\
 & & \Rightarrow S\dblortho\subseteq \Big(g^*\Big(\Big(\L(g)(S)\Big)\dblortho\Big)\Big)\dblortho=g^*\Big(\Big(\L(g)(S)\Big)\dblortho\Big) \\
 & & \Rightarrow \L(g)(S\dblortho)\subseteq \Big(\L(g)(S)\Big)\dblortho
\end{eqnarray*}
because we know that $\L(g)\dashv g^*$ (used in the first and last 
line) and continuity of $g\:G_1\parto G_2$ assures the equality in 
the above argument. So $\L(g)$ is a continuous morphism of Hilbert 
lattices.
\par
Given $f\:L_1\to L_2$, a continuous morphism of Hilbert lattices, 
$\C(f)$ is precisely defined as the unique map that makes the square 
in figure \ref{35.1} commute.
\begin{figure}
$$\xymatrix@=15mm{
L_1\ar[r]^f\ar[d]_{(\ )^{\perp\perp}} & L_2\ar[d]^{(\ )^{\perp\perp}} \\
\C(L_1)\ar[r]_{\C(f)} & \C(L_2)}$$ \caption{The definition of 
$\C(f)$}\label{35.1}
\end{figure}
By continuity of $f$, its right adjoint $f^*\:L_2\to L_1$ maps 
closed elements to closed elements; thus the restriction of $f^*$ to 
closed elements provides a right adjoint to $\C(f)$, showing that 
the latter preserves suprema. It is merely an observation that, 
because $f$ sends atoms to atoms or $0$, so does $\C(f)$. Thus 
$\C(f)$ is a morphism of propositional systems.
\par
Finally, let $h\:C_1\to C_2$ be a morphism of propositional systems. 
From the proof of \ref{108} we know that a subspace $S\subseteq 
\G(C_2)$ is closed if and only if $S=\{b\in\G(C_2)\mid b\leq 
\clbigvee S\}$; so in this case
\begin{eqnarray*}
\G(h)^*(S)
 & = & \{a\in\G(C_1)\mid h(a)=0\}\cup\G(h)^{-1}(S) \\
 & = & \{a\in\G(C_1)\mid h(a)\leq\clbigvee S\}.
\end{eqnarray*}
Writing $h^*\:C_2\to C_1$ for the right adjoint to $h$, this can be 
written as
$$\G(h)^*(S)=\{a\in\G(C_1)\mid a\leq h^*(\clbigvee S)\}$$ 
which shows that $\G(h)^*(S)$ is closed, since 
$\clbigvee(\G(h)^*(S))=h^*(\clbigvee S)$ by atomisticity.
\end{proof}
Now we can conclude this section with the following result.
\begin{theorem}\label{37}
The categories $\HilbGeom$, $\HilbLat$ and $\PropSys$ are 
equivalent\index{equivalence of categories!of Hilbert geometries, 
Hilbert lattices and propositional systems}: the assignments
\begin{eqnarray*}
 & & \L\:\HilbGeom\to\HilbLat\:\Big(g\:G_1\parto G_2\Big)\mapsto\Big(\L(g)\:\L(G_1)\to\L(G_2)\Big) \\
 & & \C\:\HilbLat\to\PropSys\:\Big(f\:L_1\to L_2\Big)\mapsto\Big(\C(f)\:\C(L_1)\to\C(L_2)\Big) \\
 & & \G\:\PropSys\to\HilbGeom\:\Big(h\:C_1\to C_2\Big)\mapsto\Big(\G(h)\:\G(C_1)\to\G(C_2)\Big)
\end{eqnarray*}
are functorial, and for a Hilbert geometry $G$, a Hilbert lattice 
$L$ and a propositional system $C$ there are natural isomorphisms 
\begin{eqnarray*}
 & & \kappa_G\:G\iso\G(\C(\L(G)))\:a\mapsto\{a\}, \\
 & & \lambda_L\:L\iso\L(\G(\C(L)))\:x\mapsto\{a\in\G(L)\mid a\leq x\}, \\
 & & \mu_C\:C\iso\C(\L(\G(C)))\:x\mapsto\{a\in\G(C)\mid a\leq x\}.
\end{eqnarray*}
\end{theorem}
\begin{proof} We shall leave some verifications to the reader: the 
functoriality of $\G$, $\L$ and $\C$, and the naturality of 
$\kappa$, $\lambda$ and $\mu$. But we shall prove that the latter 
are indeed isomorphisms.
\par
Right after \ref{108} we had already remarked that 
$\G(\C(\L(G)))=\G(\L(G))$, i.e.\ $\C(\L(G))$ and $\L(G)$ have the 
{\it same} atoms and induce the {\it same} collinearity relation, so 
we already know by \ref{17} that $\kappa_G$ is an isomorphism of 
projective geometries (it was called $\alpha_G$ in \ref{17}); we 
only need to prove that $\kappa_G$ and its inverse are continuous. A 
sufficient condition thereto is that two points $a$ and $b$ of $G$ 
are orthogonal if and only if $\kappa_G(a)$ and $\kappa_G(b)$ are 
orthogonal in $\G(\C(\L(G)))$. Indeed this fact is true:
$$a\perp b\iff a\in\{b\}\ortho\iff\{a\}\subseteq\{b\}\ortho\iff\kappa_G(a)\perp\kappa_G(b).$$
\par
Similarly as above, we already know by \ref{17} that $\lambda_L$ is 
an isomorphism of projective lattices (it was called $\beta_L$ 
before). Now let $a$ be an atom of $\C(L)$ (or of $L$) and $x\in L$, 
then 
$$a\in\lambda_L(x\ortho)\iff a\leq x\ortho\stackrel{*}{\iff} x\leq a\ortho\iff\forall b\in\lambda_L(x):b\leq a\ortho\iff a\in\lambda_L(x)\ortho$$
(the equivalence in $(*)$ uses that the atom $a$ is closed and that 
$x\leq x\dblortho$). This proves that 
$\lambda_L(x\ortho)=\lambda_L(x)\ortho$ for all $x\in L$, from which 
it follows that $\lambda_L$ and its inverse are continuous morphisms 
of Hilbert lattices.
\par
Finally, it is clear that each $\mu_C(x)\subseteq\G(C)$ is a 
subspace (for the collinearity on $\G(C)$ as in \ref{108}). Like 
before we have that $\mu_C(x\ortho)=\mu_C(x)\ortho$ for $x\in C$. It 
then follows easily that any such $\mu_C(x)$ is biorthogonally 
closed in $\L(\G(C))$, so at least is $\mu_C$ a well-defined, and 
obviously order-preserving, map between ordered sets. The 
order-preserving map
$$\eta_C\:\C(\L(\G(C)))\to C\:S\mapsto\clbigvee S$$
satisfies $\eta_C\circ\mu_C=\id_C$ by atomisticity of $C$; and in 
the proof of \ref{108} we had already shown that a subspace 
$S\subseteq\G(C)$ is closed if and only if $S=\mu_C(\eta_C(S))$, so 
$\mu_C\circ\eta_C$ is the identity too. Thus we find that $\mu_C$ and 
$\eta_C$ constitute an isomorphism of lattices and hence $\mu_C$ is an 
isomorphism in $\PropSys$, with inverse $\eta_C$.
\end{proof}
\par
The natural maps $\kappa_G$, $\lambda_L$ and $\mu_C$ are actually 
{\it more} than isomorphisms in their respective categories: they 
are examples of `ortho-isomorphisms'. For completeness' sake, we 
shall very quickly make this precise. 
\begin{definition}\label{40}
  A continuous morphism $g\:G_1\parto G_2$ between Hilbert
  geom\-e\-tries is an {\bf ortho-morphism}\index{morphism!of
    projective geometries} if $a\perp b$ implies $g(a)\perp g(b)$ for
  every $a,b$ in the domain of $g$. A continuous morphism $f\:L_1\to
  L_2$ between Hilbert lattices is an {\bf
    ortho-morphism}\index{morphism!of projective
    lattices}\index{morphism!ortho-} if $f(x\ortho)\leq f(x)\ortho$
  for all $x\in L$. And a morphism $h\:C_1\to C_2$ between
  propositional systems is an {\bf ortho-morphism}\index{morphism!of
    propositional systems} if $h(x\ortho)\leq h(x)\ortho$.
\end{definition}
It obviously makes sense to consider the subcategories 
$\HilbGeom_{\perp}$, $\HilbLat_{\perp}$ and $\PropSys_{\perp}$ of 
$\HilbGeom$, $\HilbLat$ and $\PropSys$ with the same objects but 
with ortho-morphisms. An isomorphism in one of those categories is 
called a {\bf ortho-isomorphism}. It turns out that the functors in 
\ref{37} restrict to these smaller categories, and we have actually 
shown in \ref{37} that the maps $\kappa_G$, $\lambda_L$ and $\mu_C$ 
are ortho-isomorphisms: so the three categories $\HilbGeom_{\perp}$, 
$\HilbLat_{\perp}$ and $\PropSys_{\perp}$ are equivalent too. 
\par
It is also possible to consider `non-continuous ortho-morphisms' 
between Hilbert geometries $G_1$ and $G_2$, resp.\ Hilbert lattices 
$L_1$ and $L_2$: such is a morphism $g\:G_1\parto G_2$ in 
$\ProjGeom$, resp.\ $f\:L_1\to L_2$ in $\ProjLat$, satisfying the 
appropriate orthogonality condition of \ref{40}. Two equivalent 
categories are obtained, but it is not known if there is a 
third equivalent category of propositional systems, i.e.\ if there is a suitable notion of morphism between propositional systems to 
correspond with that of `non-continuous ortho-morphism' between 
Hilbert geometries, resp.\ Hilbert lattices. See [Faure and 
Fr\"olicher, 2000, 14.3] for a discussion.

\section{Irreducible components again}\label{F}
\setcounter{theorem}{0}
By ``forgetting'' about its orthogonality relation, we may view a 
Hilbert geometry $G$ as an object of $\ProjGeom$ and consider its 
decomposition in maximal irreducible subspaces, cf.\ \ref{19.8}. We 
shall show that this coproduct actually ``lives'' in $\HilbGeom$, 
but we must start off with a couple of lemmas. First an adaptation 
of \ref{19.5}: we prove that the functor $\HilbGeom\to\ProjGeom$ 
``creates'' coproducts, from which it follows that $\HilbGeom$ has 
all coproducts and that the functor preserves these.
\begin{lemma}\label{60.1}
Given a family $(G_i,l_i,\perp_i)_{i\in I}$ of Hilbert geometries, 
the coproduct of the underlying projective geometries 
$(\disju_iG_i,l)$ in $\ProjGeom$ becomes a Hilbert geometry for the 
orthogonality relation
$$a\perp b\mbox{ if either }a\in G_k,b\in G_l,k\neq l\mbox{ or }a\perp_k b\mbox{ in some }G_k,$$
and the inclusion morphisms $(s_k\:G_k\to\disju_iG_i)_{k\in I}$ 
become continuous morphisms that form a coproduct in 
$\HilbGeom$.
\end{lemma}
\begin{proof} First we verify axioms (O1--5) of \ref{100}, using the 
notations introduced there. Clearly (O1--2) are trivial. For (O3), 
if $a,b$ both belong to some $G_k$, then also $c\in G_k$ by the 
hypothesis $l(a,b,c)$; if $p\not\in G_k$ then the conclusion holds 
trivially, and if $p\in G_k$ then we may use (O3) in 
$(G_k,l_k,\perp_k)$. If $a,b$ belong to different components then 
$c=a$ or $c=b$ so that (O3) is trivially satisfied. The argument for 
(O4): if $a\perp b$ then $q\defeq b$ does the job; if $a\not\perp b$ 
then $a$ and $b$ necessarily belong to the same component $G_k$ and 
we can apply (O4) to $(G_k,l_k,\perp_k)$. The verification of (O5) 
is a bit more tricky. We already know from the proof of \ref{19.5} 
that $S\subseteq\disju_iG_i$ is a subspace of $(\disju_iG_i,l)$ if 
and only if, for all $k$, $S\cap G_k$ is a subspace of $(G_k,l_k)$. 
Now an easy calculation shows that $S\ortho\cap G_k=(S\cap 
G_k)^{\perp_k}$, whence $S\ortho=\disju_i(S\ortho\cap 
G_i)=\disju_i(S\cap G_i)^{\perp_i}$ and in particular 
$S\dblortho=\disju_i(S\cap G_i)^{\perp_i\perp_i}$. From this it 
follows that $S$ is biorthogonally closed in $G$ if and only if, for 
all $k\in I$, $S\cap G_k$ is biorthogonally closed in $G_k$. In this 
case we can thus compute that:
\begin{eqnarray*}
S\vee S\ortho
 & = & \Big(\disju_i(S\cap G_i)\Big)\vee\Big(\disju_i(S\cap G_i)^{\perp_i}\Big) \\
 & \stackrel{*}{=} & \disju_i\Big((S\cap G_i)\vee(S\cap G_i)^{\perp_i}\Big) \\
 & = & \disju_iG_i \\
 & = & G.
\end{eqnarray*}
The equation $(*)$ follows because the coproduct $\L(\disju_i 
G_i)\cong\times_i\L(G_i)$ has the componentwise order.
\par
Since $S\subseteq G$ is closed if and only if each $S\cap G_k$ is 
closed, it is clear that each $s_k\:G_k\to\disju_iG_i$ as in 
(\ref{19.7}) is continuous: because $s_k^*(S)=S\cap G_i$. And for 
the same reason, the unique factorization of a family of {\it 
continuous} morphisms $(g_k\:G_k\parto G)$ is easily seen to be 
continuous too: if $S\subseteq G$ is closed, then each $g^*(S)\cap 
G_k=g_k^*(S)\subseteq G_k$ is closed, so 
$g^*(S)=\disju_ig_i^*(S)\subseteq\disju_iG_i$ is closed.
\end{proof}
\par
Recall that a subspace $S\subseteq G$ of a projective geometry is 
always a projective geometry in its own right (for the inherited 
collinearity), and that the inclusion $S\to G$ is always a morphism 
of projective geometries (with empty kernel). In the case of a 
Hilbert geometry $G$ we can prove that the closure by 
biorthocomplement on $G$ can be ``relativized'' to any of its {\it 
closed} subspaces.
\begin{lemma}\label{60.2}
If $S\subseteq G$ is a closed subspace of a Hilbert 
geometry\index{subspace!closed}, then $S$ is a Hilbert geometry for 
the inherited collinearity and orthogonality, and the inclusion 
$S\to G$ is a continuous morphism of Hilbert geometries.
\end{lemma}
\begin{proof} In this proof we shall use the following notation: if 
$T\subseteq S\subseteq G$ are subspaces with $S$ closed, we put 
$T'\defeq T\ortho\cap S$. We can prove a little trick: 
\begin{equation}\label{60.2.1}
 T''=(T\ortho\cap S)\ortho\cap 
S\stackrel{*}{=}(T\dblortho\vee S\ortho)\cap S 
\stackrel{**}{=}T\dblortho\vee(S\ortho\cap S)=T\dblortho.
\end{equation}
We used $S=S\dblortho$ and \ref{104} (\ref{f}) in ($*$), and 
modularity of $\L(G)$ (or orthomodularity of $\C(\L(G))$, if one 
wishes) in ($**$).
\par
The verification of (O1--4) for the projective geometry $S$ with 
inherited orthogonality is entirely straightforward. As for (O5), 
let $T\subseteq S$ be a subspace. Then (\ref{60.2.1}) says that 
$T''=T$ implies $T\dblortho=T$ which implies $T\vee T\ortho=G$, so 
that by modularity of $\L(G)$, 
$$T\vee T'=T\vee(T\ortho\cap S)=(T\vee 
T\ortho)\cap S=S.$$ 
\par
To verify the continuity of the morphism of 
projective geometries $S\hookrightarrow G$, let $T\subseteq G$ be a 
closed subspace. The intersection of closed subspaces is always a 
closed subspace, so $s^*(T)=T\cap S$ is closed in $G$. But by 
(\ref{60.2.1}) this is the same as being closed in $S$.
\end{proof}
S. Holland [1995, 3.3] explains how this lemma can be seen as 
motivation for axiom (O5) in the definition of `Hilbert geometry'.
\par
Finally we show a (remarkably strong) converse to \ref{60.1}.
\begin{lemma}\label{y0}
Given a family $(G_i,l_i)_{i\in I}$ of projective geometries, if 
their coproduct $(\disju_iG_i,l)$ in $\ProjGeom$ is in fact a 
Hilbert geometry for some orthogonality relation $\perp$, then each 
$(G_k,l_k)$ becomes a Hilbert geometry for the induced 
orthogonality, and the inclusion morphisms 
$(s_k\:G_k\to\disju_iG_i)_{k\in I}$ become continuous morphisms that 
form a coproduct in $\HilbGeom$.
\end{lemma}
\begin{proof} By the above lemmas we only need to prove that the $(G_i)_i$ 
form a pairwise orthogonal family of closed subspaces of 
$\disju_iG_i$. But let $a\in G_j$ and $b\in G_k$ for some $j\neq k$, 
then, by the coproduct construction in $\ProjGeom$, the line $a\*b$ 
can only contain the points $a$ and $b$, which thus by (O4) of 
\ref{100} must be orthogonal in the Hilbert geometry $\disju_iG_i$. 
From this and (O1) it is then easily seen that 
$G_j\ortho=\disju_{i\neq j}G_i$, so that $G_j\dblortho=G_j$ follows. 
\end{proof}
\par
Having these technical results, we shall come to the point: we begin 
with a corollary of \ref{19.8} and the above lemmas.
\begin{theorem}\label{y1}
A Hilbert geometry $G$ is the coproduct\index{subspace!maximal 
irreducible}\index{Hilbert geometry!coproduct decomposition of} in 
$\HilbGeom$ of its maximal irreducible subspaces.
\end{theorem}
This now allows us to turn the elementary notion of `irreducibility' 
for a Hilbert geometry into a categorical one by refining the result 
given in \ref{19.7.1} for more general projective geometries.
\begin{corollary}\label{y2}
A Hilbert geometry $G$ is irreducible\index{Hilbert 
geometry!irreducible} (in the sense of \ref{19.1} or equivalently 
\ref{19.7.1}) if and only if it is not a coproduct in $\HilbGeom$ of 
two (or more) non-empty Hilbert geometries.
\end{corollary}
\begin{proof} The coproduct-decompositions of a 
Hilbert geometry $(G,l,\perp)$ in $\HilbGeom$ correspond to those of 
the underlying projective geometry $(G,l)$ in $\ProjGeom$, by 
\ref{60.1} and \ref{y0}.
\end{proof}
What the above really says, is that there is only one meaning for 
the term `irreducible Hilbert geometry' $G$, namely: all projective 
lines in $G$ have at least three points, or equivalently: $G$ is not 
coproduct-decomposable in $\ProjGeom$, or equivalently: $G$ is not 
coproduct-decomposable in $\HilbGeom$. 
\par
It is a matter of exploiting the categorical equivalence of Hilbert 
geometries and Hilbert lattices to deduce the following statements 
from the above and the results in section \ref{C}.
\begin{lemma}\label{y3.0}
The category $\HilbLat$ has coproducts. Explicitly, if $(L_i)_{i\in 
I}$ are Hilbert lattices with respective orthogonality operators 
$x\mapsto x^{\perp_i}$, then the coproduct $\times_iL_i$ in 
$\ProjLat$ becomes a Hilbert lattice for the orthogonality operator
$$(x_i)_i\mapsto(x_i^{\perp_i})_i,$$
and the inclusion morphisms $s_k\:L_k\to\times_iL_i$ become 
continuous morphisms that form a universal cocone in $\HilbLat$.
\end{lemma}
\begin{proposition}\label{y3}
A Hilbert lattice $L$ is irreducible\index{Hilbert 
lattice!irreducible} (in the sense of \ref{x1}) if and only if it is 
not a coproduct in $\HilbLat$ of two (or more) non-trivial Hilbert 
lattices.
\end{proposition}
\begin{proposition}\label{y3.1} 
Let $G$ be a Hilbert geometry and $L$ a Hilbert lattice that 
correspond to each other under the equivalence 
$\HilbGeom\simeq\HilbLat$. Then $L$ is irreducible if and only if 
$G$ is.
\end{proposition}
\begin{theorem}\label{y4}
Every Hilbert lattice $L$ is a coproduct in $\HilbLat$ of 
irreducible Hilbert lattices\index{Hilbert lattice!coproduct 
decomposition of}.
\end{theorem}
\par
Finally we can state everything in terms of propositional systems; 
surely the reader is by now familiar with our way of doing this, so 
we shall once again skip all the details.
\begin{lemma}\label{y5.1}
The category $\PropSys$ has coproducts. In fact, given a family 
$(C_i)_{i\in I}$ of propositional systems with respective 
orthogonality operators $x\mapsto x^{\perp_i}$, the cartesian 
product $\times_iC_i$ with componentwise structure and the map
$$\times_iC_i\to\times_iC_i\:(x_i)_i\mapsto(x_i^{\perp_i})_i$$
as orthocomplementation, is a propositional system; and the maps
$$s_k\:C_k\to\times_iC_i\:x\mapsto(x_i)_i$$
where $x_k=x$ and $x_i=0$ if $i\neq k$, form a universal cocone in 
$\PropSys$.
\end{lemma}
\begin{definition}\label{y5}
A propositional system $C$ is {\bf irreducible}\index{propositional 
system!irreducible} if it is not a coproduct in $\PropSys$ of two 
(or more) non-trivial propositional systems.
\end{definition}
\begin{proposition}\label{y6}
Let $G$ be a Hilbert geometry, $L$ a Hilbert lattice and $C$ a 
propositional system that correspond to each other under the triple 
equivalence $\HilbGeom\simeq\HilbLat\simeq\PropSys$. Then $G$ is an 
irreducible Hilbert geometry if and only if $L$ is an irreducible 
Hilbert lattice, if and only if $C$ is an irreducible propositional 
system.
\end{proposition}
\begin{theorem}\label{y7}
Every propositional system $C$ is the coproduct in $\PropSys$ of 
irreducible propositional systems\index{propositional 
system!coproduct decomposition of}.
\end{theorem}
\par
The remark that we have made at the end of section \ref{C}, can be 
repeated here: on the one hand can the statements in \ref{y4} and 
\ref{y7} be made more precise by saying exactly which are the 
``irreducible components'' of a Hilbert lattice, resp.\ 
propositional system; on the other hand are these theorems 
particular cases of a more general principle involving `central 
elements' of lattices. Again we refer to \ref{85.1} for a comment on 
this matter.

\section{The Representation Theorem for propositional systems}\label{G}
\setcounter{theorem}{0}
This beautiful result is due to C.~Piron [1964, 1976], who 
generalized the finite-dimensional version of G.~Birkhoff and 
J.~von~Neumann [1936]. 
\begin{theorem}[Piron's Representation Theorem]\label{71}
Every irreducible propositional system of rank at least $4$ is 
ortho-isomorphic to the lattice of (biorthogonally) closed subspaces 
of a generalized Hilbert space.\index{propositional system!Piron's 
Representation Theorem for}
\end{theorem}
We can obviously put this theorem in geometric terms.
\begin{theorem}\label{72}  
For every arguesian Hilbert geometry $(G,\perp)$ there exists a 
generalized Hilbert space $(H,K,{\ }^*, \inprod{\ ,\ })$ such that 
$(G,\perp)$ is ortho-isomorphic to $(\P(H),\perp)$, where $\P(H)$ is 
given the orthogonality relation induced by $\inprod{\ ,\ 
}$.\index{Hilbert geometry!representation for arguesian}
\end{theorem} 
Moreover, the Hermitian form\index{Hermitian form} of which this 
theorem speaks, is essentially unique. 
\begin{proposition}[Uniqueness of Hermitian form]\label{73} 
  Let $H$ be a vector space over $K$. Let $\alpha\mapsto\alpha^*$ and $\alpha\mapsto\alpha^{\#}$ be
  two involutions on $K$, let $(x,y)\mapsto\inprod{x,y}$ be a $*$-Hermitian form
  and $(x,y)\mapsto[x,y]$ a $\#$-Hermitian form on $H$. If both Hermitian forms
  induce the same orthogonality on $\P(H)$, then there exists $0\neq
  \lambda=\lambda^*\in K$ such that $\rho^{\#} = \lambda^{-1}
  \rho^* \lambda$ for all $\rho \in K$ and $[x,y] = \inprod{x,y}\lambda$ for all $x,y\in H$.
\end{proposition}
The existence and uniqueness of a vector space with an anisotropic
Hermitian form inducing the Hilbert geometry $(G,\perp)$ only 
require axioms (O1) through (O4) of the definition of a Hilbert 
geometry (cf.\ \ref{100}). Axiom (O5) makes the Hermitian form 
orthomodular (see \ref{83} and \ref{82.1} for related comments). The 
field $K$ in Theorem~\ref{72} cannot be finite [Eckmann and Zabey, 
1969; Ivert and Sj\"odin, 1978]; see also [Faure and Fr\"olicher, 
2000, 14.1.12].
\par
We shall now present S. Holland's [1995, \S3] proof of the geometric 
version of Piron's theorem: it is essentially a smart application of 
the Second Fundamental Theorem to the isomorphism induced by the 
orthogonality between $G$ and its {\bf opposite geometry} $G\op$. 
The latter\index{projective geometry!opposite of} is by definition 
$\G(C\op)$, where $C\op$ is the propositional system opposite to 
$C\defeq\C(\L(G))$ and it is the isomorphism $C\cong C\op$ given by 
the orthocomplementation which induces the isomorphism $G \cong 
G\op$. Geometrically, $G\op$ has the closed 
hyperplanes\index{hyperplane} $\{p^{\perp}\mid p\in G\}$ as its 
points and its collinearity satisfies, for $p,q,r\in G$,
\begin{equation}\label{73.1}
l(p^{\perp},q^{\perp},r^{\perp})\mbox{ in }G\op\iff l(p,q,r)\mbox{ in }G \iff q=r\mbox{ or }p^{\perp}\supseteq q^{\perp}\cap r^{\perp}\mbox{ in }\L(G).
\end{equation}
\par
Given an arguesian Hilbert geometry $G$, we can by the First 
Fundamental Theorem assume that $G=\P(H)$ for a $K$-vector space $H$. 
The dual 
$$H^*\defeq\{f\:V\to K\mid f\mbox{ linear}\}$$ 
is then a right $K$-vector space, thus a left vector space over the opposite field 
$K\op$ (in which the multiplication, written with a 
centered dot, is reversed: $\rho\cdot\lambda=\lambda\rho$). For every closed hyperplane $M=(K x)^\perp$ of $\P(H)$ there 
exists a linear functional $f_x\in H^*$, unique up to scalar 
multiple, which has $M$ as its kernel. Now let $F$ be the subspace of 
$H^*$ spanned by $\{f_x\mid x \in H\}$.
\begin{lemma}\label{74}
  The map $\phi\:\P(H)\to\P(F)\: Kx\mapsto K\op\cdot f_x$
  is an isomorphism of projective geometries.
\end{lemma}
\begin{sketchofproof} With (\ref{73.1}) the lemma is proved by
  verifying that, for linear functionals $f,g,h\in H^*$, $f$ belongs
  to the $K\op$-span of $g$ and $h$ if and only if
  $\ker(f)\supset\ker(g)\cap\ker(h)$; which is an exercise in linear
  algebra [Faure and Fr\"olicher, 2000, 11.1.11].
\end{sketchofproof}
Applying the Second Fundamental Theorem to the isomorphism $\phi$, there exists a field isomorphism 
$\sigma\: K\to K\op$ and a bijective $\sigma$-linear map $A\: H 
\to F$ such that $\phi(K x) = K\op\cdot A(x)$ for all $x \in 
H$. Note that $\ker(A(y))=(K y)^{\perp}$.
\begin{lemma}\label{75}
The map $[\ ,\ ]\:H\times H \to K\:(x,y)\mapsto[x,y]\defeq A(y)(x)$ is {\bf sesquilinear}: it is additive in both factors and satisfies for all $x,y \in H$ and $\lambda \in K$
\begin{enumerate}
%
%
\setlength{\itemsep}{0pt} \setlength{\parskip}{0pt} 
\item[(Q1)] $[\lambda x,y] = \lambda[x,y]$, 
\item[(Q2)] $[x, \lambda y] = [x,y]\sigma(\lambda)$. 
\end{enumerate}
Moreover, the orthogonality $\perp$ on $G=\P(H)$ corresponds to the 
one induced by $[\ ,\ ]$, that is, $K x \perp K y \Leftrightarrow 
[x,y]=0$. Consequently, $[\ ,\ ]$ is anisotropic (i.e.\ it satisfies 
(S4) right after \ref{ghs}).
\end{lemma}
All that is left is to find the involution on $K$ and to rescale 
$[\ ,\ ]$ to make it Hermitian. Because $[\ ,\ 
]$ is anisotropic we can choose a $z \in H$ such that $\varepsilon \defeq [z,z] 
\neq 0$. Define another sesquilinear form on $H$ by putting 
$\inprod{\ ,\ } \defeq [\ ,\ ]\varepsilon^{-1}$ and set 
$\rho^*\defeq \varepsilon\sigma(\rho)\varepsilon^{-1}$ for all $\rho 
\in K$. Then $\inprod{\ ,\ }$ induces the same 
orthogonality as $[\ ,\ ]$ and it still 
satisfies (Q1) and (Q2) with $\sigma$ replaced by the 
anti-automorphism $\rho\mapsto\rho^*$ of $K$. To satisfy all 
requirements for $(H,K,{\ }^*,\inprod{\ ,\ })$ to be a 
generalized Hilbert space (cf.~\ref{ghs}) we now only need to 
prove a last result.
\begin{lemma}\label{76}
For all $x, y \in H$ and all $\rho \in K$, we have $\inprod{x,y} = 
\inprod{y,x}^*$ and $\rho^{**}=\rho$.
\end{lemma}
\begin{proof} Apply \ref{27.1} to the two maps $H \to F$ given by $y 
\mapsto \inprod{\cdot ,y}$ and $y \mapsto \inprod{y, \cdot}^{\#}$, 
where $\alpha\mapsto\alpha^{\#}$ is the inverse automorphism of $\alpha\mapsto\alpha^*$. Remembering that $F$ 
is a $K\op$-vector space, we obtain a nonzero $\xi\in K$ such 
that $\inprod{x,y}^*=\xi^*\inprod{y,x}$ for all $x,y\in H$. Then 
$1=\inprod{z,z}^*=\xi^*\inprod{z,z}=\xi^*$ because $\inprod{z,z}=1$. 
Moreover $\rho=\inprod{\rho z,z}=\inprod{z,\rho z}^*=
\inprod{\rho z,z}^{**}=\rho^{**}\inprod{z,z}^{**}=\rho^{**}$, for every $\rho \in K$,
proving that $\alpha\mapsto\alpha^*$ is an involution on $K$.
\end{proof}
This finishes the proof of \ref{71} and \ref{72}. 
Uniqueness of the Hermitian form up to scaling (as stated in \ref{73}) is 
obtained by an application of \ref{27.1}.
\par
In [Faure and Fr\"olicher, 2000, \S14.3] the reader can find
representation theorems for morphisms between projective geometries
preserving orthogonality, among which Wigner's theorem which
geometrically characterizes isometries for real, complex or
hamiltonian Hilbert spaces. See also [Faure, 2002]. We state here a
slightly more general but also well-known version.

\begin{definition}\label{semiunitary} 
  Let $H_1$ and $H_2$ be orthomodular spaces over a field $K$. An
  isomorphism $f\:H_1 \to H_2$ in $\SemiVec$ is called a {\bf
    semi-unitary}\index{semi-unitary map} if there exists a $\lambda \in K$ such that for
  all $x,y \in H_1$, we have
  $\inprod{f(x),f(y)}=\sigma(\inprod{x,y})\lambda$ where $\sigma\:K\to
  K$ is the automorphism associated to $f$. Moreover, $f$ is called
  {\bf unitary}\index{unitary map} when $\inprod{f(x),f(y)}=\inprod{x,y}$ for all $x,y\in H_1$.
\end{definition}

\begin{theorem}[Wigner]\label{wigner}\index{Wigner's Theorem}
  Let $H_1$ and $H_2$ be orthomodular spaces of dimension at least $3$
  over a field $K$. Then every ortho-isomorphism $\C(H_1) \to \C(H_2)$
  is induced by a semi-unitary map $H_1 \to H_2$.
\end{theorem}

\section{From here on}\label{H}
\setcounter{theorem}{0}
We shall end with some comments on the material that we presented in 
this chapter, and with some hints for further study.
\begin{mycomment}\label{81}
{\bf Projective closure.}\index{projective closure} In a remark 
following \ref{6.1} we have hinted at the fact that to any 
sub\textit{set} $A\subseteq G$ of a projective geometry we can 
associate the smallest sub\textit{space} $\cl(A)\subseteq G$ that 
contains $A$. It is easily verified that this operation 
$A\mapsto\cl(A)$ satisfies the following conditions:
\begin{enumerate}\label{81.0}
%
%
\setlength{\itemsep}{0pt} \setlength{\parskip}{0pt} 
\item\label{c1} it is monotone and satisfies $\cl(\cl(A))\subseteq\cl(A)\supseteq A$, 
\item\label{c2} $a\in\cl(A)$ implies $a\in\cl(B)$ for some finite subset 
$B\subseteq A$,
\item\label{c3} $x\not\in\cl(A)$ and $x\in\cl(A\cup\{b\})$ imply 
$b\in\cl(A\cup\{x\})$,
\item\label{c4} $\cl(\emptyset)=\emptyset$ and $\cl(\{a\})=\{a\}$ for all $a\in 
G$,
\item\label{c5} for non-empty $A,B\subseteq G$, $\cl(A\cup 
B)=\bigcup\{\cl(\{a,b\})\mid a\in\cl(A),b\in\cl(B)\}$.
\end{enumerate}
A set $G$ together with an operation $\cl\:2^G\to 2^G$ satisfying 
(\ref{c1}) is called a {\bf closure space}; if on top of that it 
satisfies (\ref{c2}--\ref{c3}) then it is a {\bf matroid}. A closure 
space that also satisfies (\ref{c4}) is a {\bf simple closure 
space}; and a simple matroid is often called a {\bf geometry}. If 
$A\mapsto\cl(A)$ satisfies the whole lot (\ref{c1}--\ref{c5}) then 
it is a {\bf projective closure}, and one can prove that any 
projective closure space $(G,\cl)$ is necessarily provided by a 
projective geometry. That is to say, there is an equivalence of 
categories\index{equivalence of categories!of projective geometries 
and projective closure spaces} $\ProjGeom\simeq\mathsf{ProjClos}$ of 
projective geometries on the one hand and projective closure spaces 
on the other (with appropriate morphisms). But also the `weaker' 
structures (matroids, geometries) are interesting in their own 
right; in particular can a whole deal of ``dimension theory'' for 
projective geometries (cf.\ \ref{20}) be carried out for structures 
as basic as matroids. This is the subject of Cl.-A. Faure and A. 
Fr\"olicher's [1996], see also their [2000, chapters 3 and 4].
\end{mycomment}
\begin{mycomment}\label{82}
{\bf State spaces and property lattices.} In the definition 
\ref{100} of Hilbert geometry, it follows from (O1--4) that a Hilbert
geometry is a {\bf state space}\index{state space} in the sense of
[Moore, 1995]: if $a\neq b$ then $l(q,a,b)$ and $q\perp a$ for some
$q\in G$ by (O4), but would $q\perp b$ as well then $q\perp q$ by
(O2--3) (and the symmetry of $l$) which is excluded by (O1).  That is
to say, the relation $\perp$ is irreflexive, symmetric and separating
(in the sense that $a\neq b$ implies the existence of some $q$ such
that $a\perp q\not\perp b$). Moore [1995] proves that the
biorthogonally closed subspaces of a state space form a so-called {\bf
  property lattice}\index{property lattice}: a complete, atomistic and
orthocomplemented lattice. Of course, a propositional system (cf.\
\ref{106}) is a particular example of such a `property lattice'. More
precisely, state spaces and property lattices are the objects of
equivalent categories\index{equivalence of categories!of state spaces
  and property lattices} $\sf State$ and $\sf Prop$ of which the
equivalence of $\HilbGeom$ with $\PropSys$ is a restriction. For the
relevance of $\sf State$ and $\sf Prop$ in theoretical physics see
[Moore, 1999].
\end{mycomment}
\begin{mycomment}\label{82.0}
{\bf Fewer axioms for geometries with orthogonality.} F. Buekenhout 
[1993] explains how A. Parmentier and he showed that, remarkably, (G3) of 
\ref{1} is automatically true for a set $G$ with a collinearity $l$ 
satisfying just (G1--2) and an orthogonality $\perp$ satisfying (cf.\ \ref{100})
\begin{enumerate}
%
%
\setlength{\itemsep}{0pt} \setlength{\parskip}{0pt} 
\item[(O2)] if $a\perp b$ then $b\perp a$,
\item[(O3)] if $a\neq b$, $a\perp p$, $b\perp p$ and $c \in a\*b$ then $c\perp p$,
\item[(O6)] if $a,b,p\in G$ and $a\neq b$ then there is a $q\in a\*b$ with $q\perp p$,
\item[(O7)] for all $a\in G$ there is a $b\in G$ with $a\not\perp b$.
\end{enumerate}
Clearly, (O1) implies (O7), and in the proof of \ref{102} we have 
shown that (O6) too is valid in any Hilbert geometry.
\end{mycomment}
\begin{mycomment}\label{83}
{\bf Geometries ``with extra structure''.} A Hilbert 
geometry\index{Hilbert geometry} is, by \ref{100}, a projective 
geometry with extra structure---a lot of extra structure, actually. 
There are many notions of `projective geometry with extra structure' 
that are weaker than Hilbert geometries but still have many 
interesting properties. A large part of [Faure and Fr\"olicher, 
2000] is devoted to the study of such things as {\bf Mackey 
geometries}, {\bf regular Mackey geometries}, {\bf orthogeometries} 
and {\bf pure orthogeometries}: structures that lie between 
projective geometries and Hilbert geometries. Several of these 
`geometries with extra structure' can be represented by appropriate 
`vector spaces with extra structure.' In that spirit, [Holland, 
1995, 3.6] and [Faure and Fr\"olicher, 2000, 14.1.8] are slight 
generalizations of Piron's representation theorem (cf.~\ref{71} and 
\ref{72}) which include skew-symmetric forms. 
\end{mycomment}
\begin{mycomment}\label{82.1}
{\bf On orthomodularity.} Given a vector space $V$ with an anisotropic
Hermitian form\index{Hermitian form} (i.e.\ a form satisfying (S1, S2
and S4) in \ref{ghs}) and induced orthogonality $\perp$ we have that
$(\P(V),\perp)$ satisfies (O5) if and only if $\L(V)$
satisfies (H5) if and only if the lattice of closed subspace $\C(V)$
is orthomodular if and only if the Hermitian form satisfies (S3). In
other words, and this is a key insight of Piron's [1964],
orthomodularity of $\C(V)$ is what distinguishes the generalized
Hilbert spaces among the (anisotropic) Hermitian spaces. This is one
of the reasons why orthomodular lattices\index{modular lattice!ortho-}
have been heavily studied; the standard reference on the subject is
[Kalmbach, 1983].
\end{mycomment}
\begin{mycomment}\label{82.2}
{\bf Projectors.}\index{projector}\index{Hilbert geometry!projector 
of} For a projective geometry $G$ together with a binary relation 
$\perp$ on $G$ that satisfies (O1--4) in \ref{100}, a (necessarily 
closed, cf\ \ref{104}) subspace $S\subseteq G$ satisfies $S\vee 
S\ortho=G$ if and only if for every $a\in G\setminus S\ortho$ the 
subspace $(\{a\}\vee S\ortho)\cap S$ is non-empty. In this case, 
$(\{a\}\vee S\ortho)\cap S$ is a singleton, and writing $r(a)$ for 
its single element gives a partial map 
$$r\:G\parto S\:a\mapsto r(a)\mbox{ with kernel }S\ortho$$ 
which is a retract to the inclusion $i\:S\to G$.
\begin{proof} Suppose that $S\vee S\ortho=G$ and that $a\in G\setminus (S\cup 
S\ortho)$ (if $a\in S$ then all is trivial). By the projective law, 
$a\in x\*y$ for some $x\in S$ and $y\in S\ortho$, whence $x\in a\* 
y\subseteq \{a\}\vee S\ortho$, so $x\in(\{a\}\vee S\ortho)\cap 
S\neq\emptyset$. Conversely, suppose that $a\in G\setminus(S\cup 
S\ortho)$. Pick any $x\in(\{a\}\vee S\ortho)\cap S$: thus $x\in S$ 
and $x\in a\* y$ for some $y\in S\ortho$, whence $a\in x\*y\subseteq 
S\vee S\ortho$ (using the projective law twice), which proves that 
$S\vee S\ortho=G$.
\par
Would $x_1,x_2$ be different elements of $(\{a\}\vee S\ortho)\cap S$ 
for some $a\not\in S\ortho$, then $x_1,x_2\in S$ and there exist 
$y_1,y_2\in\{a\}\vee S\ortho$ such that $x_i\in a\* y_i$ for 
$i=1,2$. Because $a\not\in S\ortho$, $a$ is necessarily different 
from the $y_i$'s. But $a$ is also different from the $x_i$'s: if 
$a=x_1$ for example, then $x_2\in x_1\* y_2$ from which $y_2\in 
x_1\* x_2\subseteq S$, which is impossible since $S\cap 
S\ortho=\emptyset$ by (O1). So we can equivalently write that $a\in 
(x_1\* y_1)\cap(x_2\* y_2)$; and by axiom (G3) of \ref{1} we get a 
point $b\in (x_1\* x_2)\cap(y_1\* y_2)$. But such $b$ lies in both 
$S$ and $S\ortho$, which is impossible. Hence the non-empty set 
$(\{a\}\vee S\ortho)\cap S$ is a singleton. In particular does this 
argument imply that $\{a\}=(\{a\}\vee S\ortho)\cap S$ if $a\in S$: 
so the partial map $r\:G\parto S$ sending $a\not\in S\ortho$ to the 
single element of $(\{a\}\vee S\ortho)\cap S$ is a retract to the 
inclusion $i\:S\to G$.
\end{proof}
Now $i\:S\to G$ is a morphism of projective geometries when we let 
$S$ inherit the collinearity from $G$, but in fact so is $r\:G\parto 
S$. Therefore ${\sf pr}\defeq i\circ r\:G\parto G$ is an idempotent 
morphism of projective geometries with kernel $S\ortho$ and image 
$S$. It is moreover true that ${\sf pr}(a)\perp b\iff a\perp{\sf 
pr}(b)$ for $a,b\not\in S\ortho$ (the morphism is ``self-adjoint''), 
and so we have every reason to speak of the {\bf projector} with 
image $S$ and kernel $S\ortho$. Much more on this can be found in 
[Faure and Fr\"olicher, 2000, section 14.4].
\end{mycomment}
\begin{mycomment} {\bf More on projectors.}\index{projector}
Interestingly, there is a lattice-theoretic analog of \ref{82.2}: A 
complete orthocomplemented lattice $C$ is orthomodular if and only 
if for each $x\in C$ the map $\phi_x\:C\to C\:y\mapsto 
x\wedge(x\ortho\vee y)$ has a right adjoint, which then is the map 
$\psi_x\:C\to C\:y\mapsto x\ortho\vee(x\wedge y)$. If $C$ is 
moreover atomistic and satisfies the covering law, then $\phi_x$ is 
a morphism of propositional systems.
\begin{proof}
Clearly the maps $\phi_x$ and $\psi_x$ preserve order. Now let C be 
a complete orthocomplemented orthomodular lattice, then 
$$\psi_x(\phi_x(y))=x\ortho\vee\Big(x\wedge\big(x\wedge(x\ortho\vee y)\big)\Big)=x\ortho\vee\Big(x\wedge(x\ortho\vee y)\Big)\stackrel{*}{=}x\ortho\vee y\geq y$$
where orthomodularity was used in $(*)$. Similarly one shows 
$\phi_x(\psi_x(y))\leq y$, so we get the adjunction 
$\phi_x\dashv\psi_x$. Conversely, if $x\leq y$ in a complete 
orthocomplemented lattice $C$, then using this information in $(**)$ 
gives
\begin{eqnarray*}
 & & \phi_{x\ortho}(y)=x\ortho\wedge(x\dblortho\vee 
y)=x\ortho\wedge(x\vee y)\stackrel{**}{=}x\ortho\wedge y\leq y,\\
 & & \psi_{x\ortho}(y)=x\dblortho\vee(x\ortho\wedge y)=x\vee(x\ortho\wedge 
y)\stackrel{**}{\leq}y\vee y=y. 
\end{eqnarray*} 
Assuming that $\phi_{x\ortho}\dashv\psi_{x\ortho}$ we get 
$y\leq\psi_{x\ortho}(y)$ from the first line, thus 
$y=\psi_{x\ortho}(y)$ if we combine it with the second line, which 
is the orthomodular law.
\par
Next suppose that $C$ is a propositional system, let $a\in C$ be an 
atom and $x\in C$. If $a\leq x\ortho$ then $\phi_x(a)=0$. If 
$a\not\leq x\ortho$ then $a\wedge x\ortho=0$ so $x\ortho\cov a\vee 
x\ortho$. By lower semimodularity of $C$ (see \ref{108} and use 
[Faure and Fr\"olicher, 2000, 1.5.7]) it follows that either 
$x\wedge x\ortho= x\wedge(a\vee x\ortho)$ or $x\wedge x\ortho\cov 
x\wedge(a\vee x\ortho)$; in any case we have shown that $\phi_x(a)$ 
is $0$ or covers $0$.
\end{proof}
A map like the $\phi_x\:C\to C$ in the statement above, is called a 
{\bf Sasaki projector}, and its right adjoint is a {\bf Sasaki 
hook}. These maps were introduced by U. Sasaki [1954], and 
extensively used in [Piron, 1976, 4--1] to describe 
lattice-theoretically the effect of an ``ideal measurement of the 
first kind'' on a (quantum) physical system. See also [Coecke and 
Smets, 2004] for a discussion of the (quantum logical) meaning of 
the adjunction of Sasaki projection and Sasaki hook.
\end{mycomment}
\begin{mycomment}\label{85.1}
{\bf Another irreducibility criterion.}\index{projective 
lattice!irreducible}\index{projective 
geometry!irreducible}\index{Hilbert 
lattice!irreducible}\index{Hilbert 
geometry!irreducible}\index{propositional system!irreducible} A {\bf 
bounded lattice} is, by definition, a lattice with a smallest 
element $0$ and a greatest element $1$. If $L_1$ and $L_2$ are 
bounded lattices then, with componentwise lattice structure, the 
cartesian product $L_1\times L_2$ is a bounded lattice too. An 
element $z\in L$ of a bounded lattice is {\bf central} if there 
exist bounded lattices $L_1$, $L_2$ and an isomorphism (i.e.\ a 
bijection that preserves and reflects order) $\phi\:L_1\times 
L_2\iso L$ such that $z=\phi(1,0)$. The set $\Z(L)$ of central 
elements, called the {\bf center} of $L$, is an ordered subset of 
$L$ that contains at least $0$ and $1$. A wealth of information on 
this topic can be found in [Maeda and Maeda, 1970, sections 4 and 5] 
or any other standard reference on lattice theory.
\par
One can easily figure out that the cartesian product $L_1\times L_2$ 
of bounded lattices is a projective lattice if and only if $L_1$ and 
$L_2$ are projective lattices (just view such an $L_i$ as a segment 
in $L$); and $L_1\times L_2$ is then a coproduct in $\ProjLat$ (see 
also \ref{19.10}). Hence $L$ is an irreducible projective lattice if 
and only if $\Z(L)=\{0,1\}$ (``$L$ has a trivial center''). One can 
moreover show that $\Z(L)$ forms a complete atomistic Boolean (i.e.\ 
complemented and distributive) sublattice of $L$; and the segments 
$[0,\alpha]\subseteq L$, with $\alpha$ an atom of $\Z(L)$, are 
precisely the `maximal irreducible segments' of $L$ (a notion that 
we did not bother defining in section \ref{C}); so $L$ is the 
coproduct in $\ProjLat$ of these segments. Details are in [Maeda and 
Maeda, 1970, 16.6] for example, where the term `modular matroid 
lattice' is used synonymously for `projective lattice'.
\par
For a propositional system $C$, one can work along the same lines to 
prove that $C$ is irreducible if and only if $\Z(C)=\{0,1\}$; the 
center $\Z(C)$ is again always a complete atomistic Boolean 
sublattice of $C$; and $C$ is the coproduct in $\PropSys$ of the 
segments $[0,\alpha]$ with $\alpha$ an atom of $\Z(C)$. C. Piron 
[1976, p.~29] has called the atoms of $\Z(C)$ the {\bf 
superselection rules} of the propositional system $C$. In geometric 
terms, a subspace $S\subseteq G$ of a projective geometry is a 
central element in $\L(G)$ if and only if also the set-complement 
$S\c\defeq G\setminus S$ is a subspace of $G$. And if $G$ is a 
Hilbert geometry then $S$ is central if and only if $S\c=S\ortho$, 
in which case $S$ is necessarily a closed subspace. So 
$\Z(\L(G))\cong\Z(\C(\L(G)))$ for a Hilbert geometry $G$, proving at 
once that the center of a Hilbert lattice $L$ is the same Boolean 
algebra as the center of the propositional system $\C(L)$ of closed 
elements in $L$.
\end{mycomment}
\begin{mycomment}\label{85}
{\bf Modules on a ring.} Vector spaces on fields are very particular 
examples of modules on rings; and modules on rings are very 
``categorical'' objects: consider a (not necessarily commutative) 
ring $R$ as a one-object $\sf Ab$-enriched category $\cal R$, then a 
(left) module $(M,R)$ is an $\sf Ab$-presheaf ${\cal M}\:{\cal 
R}\to{\sf Ab}$. (As usual, $\sf Ab$ denotes the category of abelian 
groups.) In the same vein, also semilinear maps\index{semilinear 
map} between vector spaces are instances of an intrinsically 
categorical notion: viewing ring-modules $(M,R)$ and $(N,S)$ as $\sf 
Ab$-presheaves ${\cal M}\:{\cal R}\to{\sf Ab}$ and ${\cal N}\:{\cal 
S}\to{\sf Ab}$, a ``semilinear map'' $(f,\sigma)\:(M,R)\to(N,S)$ 
ought to be defined as an $\sf Ab$-functor $\sigma\:{\cal R}\to{\cal 
S}$ together with and $\sf Ab$-natural transformation $f\:{\cal 
M}\tto{\cal N}\circ\sigma$, cf.\ figure \ref{888}.
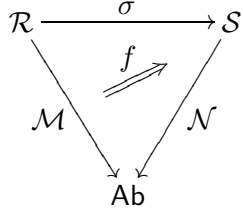
\begin{figure}
$$\xy
\xymatrix@=3mm{ & & \\ \ar@{=>}[rru]^f} \POS(-10,5) \xymatrix@=8mm{
{\cal R}\ar[rr]^{\sigma}\ar[rdd]_{{\cal M}} & & {\cal S}\ar[ldd]^{{\cal N}} \\
 & &  \\
 & {\sf Ab} & }
\endxy$$
\caption{Categorical definition of `semilinear map'} \label{888}
\end{figure}
It is thus natural to investigate whether and how one can associate 
a (suitably adapted notion of) `projective geometry' to a general 
module, and a morphism of projective geometries to a semilinear map 
as defined above. M. Greferath and S. Schmidt's Appendix E in [Gr\"atzer, 1998] and 
Faure's [2004] deal with aspects of this; in our opinion it would be enlightening to 
both algebraists and geometers to study the ($\sf Ab$-enriched) 
categorical side of this.
\end{mycomment}
\begin{mycomment}\label{86}
{\bf Lattice-theoretic equivalents to Sol\`er's condition.} Recall 
that M.P. Sol\`er [1995] proved that an infinite dimensional 
generalized Hilbert space is a ``classical'' Hilbert space (over 
$\RR, \CC$ or $\HH$) exactly when it has an orthonormal sequence 
(see [Prestel, 2006] in this volume for much more on this 
theorem).\index{generalized Hilbert space!Sol\`er's Theorem} As 
Sol\`er pointed out in the same paper, the ``angle bisecting'' axiom 
of R.P. Morash [1973] provides an equivalent but lattice-theoretic 
condition. S. Holland [1995, \S4] used ``harmonic conjugates'' to 
formulate another lattice-theoretic alternative. He also proposed 
[Holland, 1995, \S5] a (non lattice-theoretic) ``ample unitary group 
axiom''\index{generalized Hilbert space!ample unitary group axiom}: 
an infinite dimensional orthomodular space $H$ over $K$ is a 
``classical'' Hilbert space if and only if for any two orthogonal 
nonzero vectors $a,b \in H$ there exists a unitary map $U:H\to H$ 
(see \ref{semiunitary}) such that $U(Ka)=Kb$. R. Mayet [1998] has 
proved the following lattice-theoretic alternative: an orthomodular 
space $H$ is an infinite dimensional Hilbert space over $\RR, \CC$ 
or $\HH$ if and only if there exist $a,b \in \C(H)$ where $\dim~b\ge 
2$ and an ortho-isomorphism $f: \C(H) \to \C(H)$ such that 
$f|_{[0,b]}$ is the identical map and $f(a) \lneqq a$. The condition 
on $f|_{[0,b]}$ guarantees that the semi-unitary map inducing $f$ 
(by Wigner's theorem, see \ref{wigner}) is unitary. Similar 
characterizations using ``symmetries'' of the lattice $\C(H)$ were 
proposed in [Aerts and Van Steirteghem, 2000] and in [Engesser and 
Gabay, 2002]. The question whether the transitivity of the whole 
group of ortho-isomorphisms of $\C(H)$ still characterizes the 
``classical'' Hilbert spaces among the infinite dimensional 
orthomodular spaces seems to be unanswered.
\end{mycomment}

\section{Appendix: notions from lattice theory}
\setcounter{theorem}{0}
%
%
%

Mostly to fix terminology, we recall the notions from lattice theory 
we have used in this chapter; in the previous sections these are 
marked with a ``$\dagger$'' when they are used for the first time.
\par
A {\bf partially ordered set}, also called simply {\bf ordered set} or
{\bf poset}, is a set $P$ together with a binary relation $\leq$ which
is reflexive, antisymmetric and transitive. We also use the standard
notation $x < y$ for $ x \leq y \textrm{ and } x \neq y$. The {\bf
  opposite} ordered set $P\op$ has the same elements as $P$ but
with its order relation $\preccurlyeq$ reversed: for $x, y \in P$ we
have $x\leq y \iff y\preccurlyeq x$.
\par
For a
subset $X$ of $P$ we say that $p \in P$ is an {\bf upper bound} of $X$
if $x \leq p$ for every $x \in X$, we say that $p$ is a {\bf least
  upper bound}, or a {\bf supremum}, or a {\bf join}, of $X$ if for
every other upper bound $q$ we have $p \leq q$. By antisymmetry a
least upper bound is unique if it exists. The concept of {\bf
  greatest} {\bf lower bound} (also called {\bf infimum} or {\bf
  meet}) is defined dually. If the supremum of $X$ exists and lies in
$X$ we call it the {\bf maximum} of $X$, denoted by $\max X$. Dually,
we can define $\min X$, the {\bf minimum} of $X$.
\par
A {\bf lattice} is a poset $L$ any two of whose elements $x,y \in L$ 
have a {\bf meet} denoted by $x \wedge y$ and a {\bf join} denoted 
by $x \vee y$. It is {\bf complete} if any subset $X \subseteq L$ 
has a join, then denoted by $\bigvee X$, and a meet $\bigwedge X$. 
(In fact, if all joins exist in an ordered set $L$ then so do all 
meets, and {\it vice versa}; thus an ordered set $L$ is a complete 
lattice if and only if it has all joins, if and only if it has all 
meets.) Putting $X=L$ we see that a complete lattice has a {\bf 
bottom element} $0$ and a {\bf top element} $1$, that is, elements 
satisfying $0 \leq x \leq 1$ for every $x \in L$. For $a\leq b$ in a 
lattice $L$, the {\bf interval} or {\bf segment} $[a,b]$ is the 
lattice $\{x \in L \mid a\leq x\leq b\}$.
\par
A map $f\: P_1 \to P_2$ between two ordered sets is said to {\bf
  preserve order}, or is called {\bf monotone}, if for any $x,y \in
P_1$,
$$x\leq y\mbox{ implies }f(x)\leq f(y).$$  It is an {\bf isomorphism} of ordered sets 
(or of lattices when appropriate) if it moreover has an order-preserving inverse.
\par
For two elements $x,y$ of $P$ we say that $y$ {\bf covers} $x$ and we
write $x \cov y$ when $x <y$ but never $x <p <y$ for $p\in P$. If $P$
is a poset with bottom element $0$, we call $a \in P$ an {\bf atom} if
$a$ covers $0$. If $P$ is a poset with top element $1$ then $c$ is a
{\bf coatom} if $1$ covers $c$. A lattice $L$ with bottom element $0$
is called\footnote{G.  Birkhoff [1967] calls these `atomic' or `point
  lattices'.} {\bf atomistic} if every element $x\in L$ is the join of
the atoms it contains: $x=\bigvee\{a\in L\mid a\mbox{ atom, }a\leq
x\}.$
\par
A nonempty subset $D\subseteq P$ of a poset is called {\bf directed}
if for any $x,y\in D$, there exists $z \in D$ such that $x\leq z$ and
$y\leq z$. A complete lattice $L$ is called {\bf continuous} (some say
{\bf meet-continuous}) if for any directed set $D \subseteq L$ and any
$a\in L$ we have $a\wedge(\bigvee D)=\bigvee\{a\wedge d\mid d\in D\}$.
\par
A lattice $L$ is called {\bf modular} if, for every $x,y,z\in L$,
$$x\leq z\mbox{ implies }x\vee (y \wedge z) = (x\vee y)\wedge z.$$
The following are weaker notions: $L$ is {\bf upper semimodular} if 
$u \wedge v \cov v$ implies $u \cov u \vee v$; and it is {\bf lower 
semimodular} if $u \cov u \wedge v$ implies $u \wedge v \cov v$.
\par
A lattice $L$ with $0$ satisfies the {\bf covering law} if for any
$x\in L$ and any atom $a\in L$ we have
$$a\wedge x=0\mbox{ implies }x\cov a\vee x.$$
\par
An {\bf orthocomplementation} on a lattice $L$ with $0$ and $1$ is a
map $L \to L\: x \mapsto x^{\perp}$ which satisfies, for all $x,y\in
L$,
\begin{enumerate}
%
%
\setlength{\itemsep}{0pt} \setlength{\parskip}{0pt} 
\item $x\leq y$ implies $y^{\perp}\leq x^{\perp}$,
\item $(x^{\perp})^{\perp}=x$,
\item $x \vee x^{\perp}=1$ and $x \wedge x^{\perp} =0$.
\end{enumerate}
A lattice is called {\bf orthocomplemented} if it is equipped with 
an orthocomplementation. Such a lattice $L$ is called {\bf 
orthomodular} if moreover, for all $x,y\in L$,
$$x\leq y\mbox{ implies }x\vee(x^{\perp}\wedge y)=y.$$  
Since the orthocomplementation induces an isomorphism $L \to L^{\op}$
this is equivalent to
$$x\leq y\mbox{ implies }y\wedge(y^{\perp}\vee x)=x.$$
\par
Given two order-preserving maps $f\:P_1 \to P_2$ and $g\: P_2 \to P_1$
in opposite directions, we say that $f$ is a {\bf left adjoint} of
$g$, and $g$ a {\bf right adjoint} of $f$, written $f \dashv g$, if
they satisfy one, and hence all, of the following equivalent
conditions:
\begin{enumerate}
%
%
\setlength{\itemsep}{0pt} \setlength{\parskip}{0pt} 
\item\label{adj1} for all $x \in P_1$ and $y \in P_2$ we have \(f(x)
  \leq y \iff x \leq g(y)\),
\item\label{adj2} $f(x) = \min \{y \in P_2 \mid x \leq g(y)\}$ for all
  $x \in P_1$,
\item\label{adj3} $g(y) = \max \{x \in P_1 \mid f(x) \leq y\}$ for all
  $y \in P_2$,
\item\label{adj4} $x\leq g(f(x))$ and $f(g(y))\leq y$ for all $x\in
  P_1$ and all $y\in P_2$.
\end{enumerate}
The pair $(f,g)$ is called a {\bf Galois connection} or said to form
an {\bf adjunction} (between the ordered sets $P_1$ and $P_2$.)  It
follows from conditions (\ref{adj2}) and (\ref{adj3}) above that
adjoints determine each other uniquely (when they exist). And one can
check that $f$ is surjective if and only if $g$ is injective, if
and only if $f(g(y))=y$ for all $y\in P_2$.
\par
Still considering such an adjunction $f\dashv g$, $f$ preserves all
joins that exist in $P_1$; similarly, $g$ preserves all meets that
exist in $P_2$. Conversely, for an ordered set $L$ the following
conditions are equivalent:
\begin{enumerate}
%
%
\setlength{\itemsep}{0pt} \setlength{\parskip}{0pt} 
\item $L$ is a complete lattice,
\item every map $h\:L\to P$ preserving all joins has a right adjoint,
\item every map $h\:L\to Q$ preserving all meets has a left adjoint.
\end{enumerate}
\par
A {\bf closure operator} on a poset $P$ is a monotone map $\cl\:P\to
P$ satisfying, for all $x\in P$,
\begin{enumerate}
%
%
\setlength{\itemsep}{0pt} \setlength{\parskip}{0pt} 
\item $\cl(\cl(x))\leq \cl(x)$,
\item $x \leq \cl(x)$.
\end{enumerate}
It is obvious that $\cl(\cl(x))=\cl(x)$, i.e.\ that a closure operator
is an idempotent map. Its fixpoints are often said to be the {\bf
  closed elements} of $P$ (w.r.t.\ $\cl$): they form a sub-poset
$\cl(P)\subseteq P$. The surjection $\cl\: P\to\cl(P)$ and the
inclusion $i\:\cl(P)\to P$ form an adjunction $\cl\dashv i$.
Conversely, for any adjunction $f\dashv g$ between posets $P_1$ and
$P_2$, $g\circ f$ is a closure operator on $P_1$; and if moreover
$f\circ g$ is the identity on $P_2$ then $P_2$ is isomorphic to the
poset of fixpoints of $g\circ f$.
\par
One now easily deduces that, for a closure operator $\cl$ on a
complete lattice $L$, also $\cl(L)$ is a complete lattice for the 
order inherited from $L$: it has the ``same'' meets as $L$ (since 
$i$ preserves meets) and the joins are given by $\clbigvee S= 
\cl(\bigvee S)$ where $S \subseteq \cl(L)$ and $\bigvee$ is the join 
in $L$.

%
%
\index{category!of vector spaces|see{vector space}} 
\index{category!of generalized Hilbert spaces|see{Hilbert space}} 
\index{category!of projective geometries|see{projective geometry}} 
\index{category!of Hilbert geometries|see{Hilbert geometry}} 
\index{category!of projective lattices|see{projective lattice}} 
\index{category!of Hilbert lattices|see{Hilbert lattice}} 
\index{category!of propositional systems|see{propositional system}} 
\index{Hilbert space|see{generalized Hilbert space}}


\begin{thebibliography}{99}

\bibitem{aertsvs} [Diederik Aerts and Bart Van Steirteghem, 2000]
  Quantum axiomatics and a theorem of M.P.~Sol\`er, {\em Int. J.
    Theor. Phys.} {\bf 39}, pp.~497--502.

\bibitem{amemiyaaraki66} 
[Ichiro Amemiya and Huzihiro Araki, 1966] A
  Remark on Piron's Paper, {\em Publ. Res. Inst. Math. Sci. Ser. A}
  {\bf 2}, pp.~423--427.

\bibitem{artin57} 
[Emil Artin, 1957] {\em Geometric algebra}, Interscience Publishers, 
New York.

\bibitem{baer52} 
[Reinhold Baer, 1952] {\em Linear algebra and
    projective geometry}, Academic Press, New York.

\bibitem{beutelspacherrosenbaum98}
[Albrecht Beutelspacher and Ute Rosenbaum, 1998], {\em Projective 
Geometry. From foundations to applications}, Cambridge University 
Press, Cambridge.

\bibitem{birkhoff67} 
[Garret Birkhoff, 1967] {\em Lattice theory (3rd
    edition)}, Am. Math. Soc., Providence.


\bibitem{birkhoffvonneumann36} 
[Garret Birkhoff and John von Neumann,
  1936] The logic of quantum mechanics, {\em Ann. of Math.} {\bf 37}, 
  pp.~823--843.
  
\bibitem{borceux94}
[Francis Borceux, 1994] {\em Handbook of categorical algebra (3 
volumes)}, Cambridge University Press, Cambridge.


\bibitem{buekenhout93} [Francis Buekenhout, 1993] A theorem of
  {P}armentier characterizing projective spaces by polarities, {\em
    Finite geometry and combinatorics (Deinze, 1992)}, London Math.
  Soc. Lecture Note Ser. \textbf{191}, Camridge University Press, pp.~69--71.

\bibitem{coeckesmets04}
[Bob Coecke and Sonja Smets, 2004] The {S}asaki hook is not a 
[static] implicative connective but induces a backward [in time] 
dynamic one that assigns causes, {\em Internat. J. Theoret. Phys.} 
{\bf 43}, pp. 1705--1736.

\bibitem{eckmannzabey69}
[Jean-Pierre Eckmann and Ph.-Ch. Zabey, 1969] Impossibility of 
quantum mechanics in a Hilbert space over a finite field, {\em
    Helv. Phys. Acta} {\bf 42}, pp.~420--424.

\bibitem{engessgab02} [Kurt Engesser and Dov M. Gabay, 2002] Quantum
  logic, {H}ilbert space, revision theory, {\em Artificial
    Intelligence} {\bf 136}, pp.~61--100.

\bibitem{faure02} 
[Claude-Alain Faure, 2002] An elementary proof of
  the fundamental theorem of projective geometry, {\em Geom. Ded.}
    {\bf 90}, pp.~145--151.

\bibitem{faure04} 
[Claude-Alain Faure, 2004] Morphisms of projective spaces over rings,
  {\em Adv. Geom.} {\bf 4}, pp.~19--31.
  
\bibitem{faurefrolicher93} 
[Claude-Alain Faure and Alfred Fr\"olicher,
  1993] Morphisms of projective geometries and of corresponding
  lattices, {\em Geom. Ded.} {\bf 47}, pp.~25--40.

\bibitem{faurefrolicher94} 
[Claude-Alain Faure and Alfred Fr\"olicher,
  1994] Morphisms of Projective Geometries and Semilinear Maps, {\em
    Geom. Ded.} {\bf 53}, pp.~237--262.

\bibitem{faurefrolicher95} 
[Claude-Alain Faure and Alfred Fr\"olicher,
  1995] Dualities for Infinite-Dimensional Projective Geometries, {\em
    Geom. Ded.} {\bf 56}, pp.~225--236.

\bibitem{faurefrolicher96} 
[Claude-Alain Faure and Alfred Fr\"olicher,
  1996] The dimension theorem in axiomatic geometry, {\em
    Geom. Ded.} {\bf 60}, pp.~207--218.

\bibitem{faurefrolicher2000} 
[Claude-Alain Faure and Alfred
  Fr\"olicher, 2000] {\em Modern projective geometry}, Kluwer Academic
  Publishers, Dordrecht.

\bibitem{gratzer98}
[George Gr{\"a}tzer, 1998] {\em General lattice theory (2nd 
edition)}, Birkh\"auser Verlag, Basel.

\bibitem{holland95} 
[Samuel S. Holland, Jr., 1995] Orthomodularity in
  infinite dimensions; a theorem of M. Sol\`er, {\em Bull. Amer.
    Math. Soc.} {\bf 32}, pp.~205--234.
    
\bibitem{ivertsjodin1978}
[Per-Anders Ivert and Torgny Sj\"odin, 1978] On the impossibility of 
a finite propositional system for quantum mechanics, {\em
    Helv. Phys. Acta} {\bf 51}, pp.~635--636.

\bibitem{kalmbach83} 
[Gudrun Kalmbach, 1983] {\em Orthomodular lattices}, London 
Mathematical Society Monographs \textbf{18}, Academic Press Inc., 
London.

\bibitem{keller80} 
[Hans Arwed Keller, 1980] Ein nicht-klassicher Hilbertscher Raum,
{\em Math.\ Z.} {\bf 172}, pp.~41--49.

\bibitem{maclane71}
[Saunders Mac Lane, 1971] {\em Categories for the working 
mathematician}, Springer-Verlag, New York.

\bibitem{maedamaeda70} 
[Fumitomo Maeda and Shuichiro Maeda, 1970] {\em
    Theory of Symmetric Lattices}, Springer-Verlag, New York.

\bibitem{mayet98} [Ren\'e Mayet, 1998] Some characterizations of the
  underlying division ring of a Hilbert lattice by automorphisms,
  {\em Int. J. Theor. Phys.} {\bf 37}, pp.~109--114.

\bibitem{moore95} 
[David J. Moore, 1995] Representations of physical
  systems, {\em Helv. Phys. Acta} {\bf 68}, pp.~658--678.

\bibitem{moore99}
[David J. Moore, 1999] On state spaces and property lattices, {\em Stud. Hist. Philos. Sci. B Stud. Hist. Philos. Modern Phys.} {\bf 30}, pp.~61--83.

\bibitem{morash73} [Ronald P. Morash, 1973] Angle bisection and
  orthoautomorphisms in Hilbert lattices, {\em Canad. J. Math.} {\bf
    25}, pp.~261--272.

\bibitem{moulton02} 
[Forest Ray Moulton, 1902] A simple
  non-Desarguesian plane geometry, {\em Trans. Amer. Math. Soc.} {\bf
    3}, pp.~192--195

\bibitem{piron64} 
[Constantin Piron, 1964] Axiomatique Quantique, {\em
    Helv. Phys. Acta} {\bf 37}, pp.~439--468.

\bibitem{piron76} 
[Constantin Piron, 1976] {\em Foundations of Quantum Physics}, W. A. 
Benjamin, Inc., Reading.

\bibitem{prestel06}
[Alexander Prestel, 2006] Sol\`er's theorem and its history, {\em 
this volume.}

\bibitem{sasaki54}
[Usa Sasaki, 1954] Orthocomplemented lattices satisfying the 
exchange axiom, {\em J. Sci. Hiroshima Univ. Ser. A.} {\bf 17}, pp. 
293--302.

\bibitem{schwartz70}
[Laurent Schwartz, 1970] {\em Analyse, deuxi\`eme partie: Topologie 
g\'en\'erale et analyse fonctionnelle}, Hermann, Paris.
 
\bibitem{soler95} 
[Maria P. Sol\`er, 1995] Characterization of Hilbert spaces with 
orthomodular spaces, {\em Comm. Algebra} {\bf 23}, pp.~219--243.

\end{thebibliography}
\end{document}